\documentclass[10pt,journal]{IEEEtran}

\usepackage[utf8]{inputenc} % allow utf-8 input
\usepackage{float}
\usepackage{subcaption}
\usepackage[T1]{fontenc}    % use 8-bit T1 fonts
\usepackage{hyperref}       % hyperlinks
\usepackage{url}            % simple URL typesetting
\usepackage{booktabs}       % professional-quality tables
\usepackage{nicefrac}       % compact symbols for 1/2, etc.
\usepackage{microtype}      % microtypography
\usepackage{amsmath, amsfonts, graphicx}

\usepackage{cancel}
\def\x{{\mathbf x}}
\def\m{{\mathbf m}}
\def\X{{\mathbf X}}
\def\E{{\mathbb E}}
\def\V{{\mathbb V}}
\def\y{{\mathbf y}}
\def\Y{{\mathbf Y}}
\def\I{{\mathbf I}}
\def\R{{\mathbb R}}

\def\N{{\mathbb N}}
\def\cN{{\cal N}}
\def\MVN{{\text{MVN}}}
\def\lat{{\text{lat}}}
\def\obs{{\text{obs}}}
\def\st{\sigma_\mathbf{t}}
\def\sf{\sigma_\mathbf{f}}
\def\Ht{H_\mathbf{t}}
\def\Hf{H_\mathbf{f}}
\def\Mt{M_\mathbf{t}}
\def\Mf{M_\mathbf{f}}
\def\Ktr{K_\mathbf{tr}}
\def\Kti{K_\mathbf{ti}}
\def\Kr{K_\mathbf{r}}
\def\Ki{K_\mathbf{i}}
\def\er{\epsilon_\mathbf{r}}
\def\ei{\epsilon_\mathbf{i}}
\def\Wr{W_\mathbf{r}}
\def\Wi{W_\mathbf{i}}
\def\Xr{\X_\mathbf{r}}
\def\Xi{\X_\mathbf{i}}
\def\Yr{\Y_\mathbf{r}}
\def\Yi{\Y_\mathbf{i}}
\def\Wa{{\vec{W}}}
\def\Ha{{\vec{H}}}
\def\Ya{{\vec{Y}}}

\newcommand{\Normal}[1]{\mathcal{N} \left(#1\right)}

\usepackage{color}

\providecommand{\revised}[1]{{#1}}

\usepackage{amsthm}
\newtheorem{proposition}{Proposition}
\newtheorem{remark}{Remark}

\begin{document}
\title{Bayesian Reconstruction of Fourier Pairs}

\author{Felipe Tobar,
        Lerko Araya-Hernández,
        Pablo Huijse,  
        Petar M. Djuri\'c% <-this % stops a space
\thanks{F. Tobar is with the Center for Mathematical Modeling and the Department of Mathematical Engineering, Universidad de Chile. L. Araya-Hernández is with the Department of Electrical Engineering, Universidad de Chile. P. Huijse is with the Informatics Institute, Universidad Austral de Chile and the Millennium Institute of Astrophysics. P. M. Djuri\'c is with the Department of Electrical Engineering, Stony Brook University.}
}% <-this % stops a space
%\thanks{}% <-this % stops a space

\markboth{Author copy accepted at IEEE Transactions on Signal Processing (Nov.~2020)}%
{Tobar \etal: Bayesian Reconstruction of Fourier Pairs}

\maketitle

%!TEX root = ./BRFP.tex
\begin{abstract}
In a number of data-driven applications such as detection of arrhythmia, interferometry or audio compression, observations are acquired indistinctly in the time or frequency domains: temporal observations allow us to study the spectral content of signals (e.g., audio), while frequency-domain observations are used to reconstruct temporal/spatial data (e.g., MRI). Classical approaches for spectral analysis rely either on i) a discretisation of the time and frequency domains, where the fast Fourier transform stands out as the \textit{de facto} off-the-shelf resource, or ii) stringent parametric models with closed-form spectra. However, the general literature fails to cater for missing observations and noise-corrupted data. \revised{Our aim is to address the lack of a principled treatment of data acquired indistinctly in the temporal and frequency domains in a way that is robust to missing or noisy observations, and that at the same time models uncertainty effectively.} To achieve this aim, we first define a joint probabilistic model for the temporal and spectral representations of signals, to then perform a Bayesian model update in the light of observations, thus jointly reconstructing the complete (latent) time and frequency representations. The proposed model is analysed from a classical spectral analysis perspective, and its implementation is illustrated through intuitive examples. Lastly, we show that the proposed model is able to perform joint time and frequency reconstruction of real-world audio, healthcare and astronomy signals, while successfully dealing with missing data and handling uncertainty (noise) naturally \revised{against both classical and modern approaches for spectral estimation.}
\end{abstract}

%% Note that keywords are not normally used for peerreview papers.
\begin{IEEEkeywords}
Spectral estimation, Bayesian inference, Fourier pairs, interferometry 
\end{IEEEkeywords}

\IEEEpeerreviewmaketitle
%!TEX root = ./BRFP.tex

\section{Introduction}
\label{sec:intro}

\subsection{Spectral and temporal representations} % (fold)

For any given quantity that varies through time, also referred to as \emph{signal}, we can consider at least the following two representations: i) the temporal one given by the values of the signal at different time instants, and ii) the spectral representation, where the information in the signal is encoded into the activation of oscillatory components at different frequencies. Throughout the paper, we will refer to the index set for signals as \emph{time}, however, our analysis and contributions are of general interest and can be applied to spatial data, as well as other high-dimensional inputs.

In several disciplines within science and engineering, such as astronomy, brain computer interface, audio compression, finance and fault detection, one of the above representations might be preferred due to, e.g., interpretation and decision-making purposes. However, in the majority of cases, the representation cannot be \emph{chosen}, but rather it is determined by the acquisition device, meaning that the signals might be sensed in one domain but need to be analysed in a different one. For instance, in radio astronomy, interferometers such as those at ALMA (Chile) 
\cite{NRAO} observe the sky through partial sampling of its spectral representation, yet astronomers need the spatial (optical) representation to perform analyses and draw key conclusions. Conversely, in healthcare, arrhythmia can be detected by identifying the predominant frequencies of a heart-rate signal, however, the measurements are collected via a mechanical transducer that senses temporal variations of heart activity. In addition to the discrepancy between the available and sought-after representations, high quality observations---in either domain---require prohibitively expensive machinery, therefore, moving back and forth between representations becomes a clear practical challenge. 

We consider the Fourier spectrum of stationary signals, in which case the temporal and spectral representations of a signal are referred to as \emph{Fourier pairs}. As a consequence of the one-to-one property of the Fourier transform, if one of the representations (e.g., time) is completely known, then, the alternative representation (e.g., frequency) can be directly computed without uncertainty. However, in real-world scenarios such as those described above, the observations are usually partial, meaning that some values are not available, and/or are corrupted by observation noise. Therefore, and owing to the convolutional structure of the Fourier transform, if a single value of the temporal representation is missing or contaminated with observation noise, then, the entire spectral representation becomes corrupted (and vice versa).

\subsection{The classical perspective to spectrum analysis} % (fold)
\label{sub:the_classical_approach}
Methods for analysing the relationship between spectral and temporal representations are at the heart of the Signal Processing (SP) toolkit since its  dawn in the late 1940's. Spectral analysis~\cite{Stoica2005, kay:88} can be found in several flavours of SP, regardless of whether the signals under study are defined for discrete or continuous time, or whether their treatment is  statistical or deterministic. In practice, we can identify two approaches to spectral analysis, each one with their own advantages and disadvantages. First, we have nonparametric methods, such as the Welch periodogram \cite{Welch1967} which directly computes the discrete Fourier transform, or its computationally-efficient implementation, the fast Fourier transform (FFT) \cite{1965-cooley}. These methods lack structural assumptions, since they aim to be as general as possible without suffering from \emph{mismodelling}. However, under the presence of missing or corrupted data, nonparametric methods provide no basis to discriminate between true signal patterns and observation artefacts; as a consequence, these artefacts immediately propagate from observations to the spectral/temporal estimates. A second perspective to spectral estimation is that of parametric models, which incorporate known structure in the processes at hand. This enables us to jointly clean or \emph{filter} our data when observations are missing or are corrupted, and then perform spectral analysis. These parametric methods can be understood as a projection of our observations onto a space of signals with known spectral representation (such as autoregressive processes based on \cite{Yule267,Walker518}, or sum of sinusoids via the Lomb-Scargle method \cite{Lomb1976,Scargle1982}). Therefore, their analysis is computationally efficient but biased towards the chosen model space. Despite its restrictive assumptions, a key advantage of the parametric approach is that it can be readily paired with a Bayesian treatment. This concept allows us to define an observation stage to then perform spectral analysis by means of Bayesian inference. This has been pioneered by \cite{jaynes1987bayesian,bretthorst2013bayesian}, and more recently by using a Kalman  filter \cite{qi_minka_picard} or Gaussian  processes  \cite{TobarNIPS2018,protopapas}, yet  only for spectral estimation and not the \emph{inverse} task related to temporal reconstruction from spectral measurements. 
% subsection the_classical_approach (end)
 
\subsection{A novel perspective to Fourier pairs} % (fold)

\revised{The above methods are focused on one of the representations exclusively, and none of them address the \emph{Fourier-pair problem}, that is, to use observations from either domain (time or frequency) to jointly reconstruct both representations. However, simultaneously modelling both representations is critically desired for two reasons. The first one, mentioned above, is practical and stems from the fact that the acquisition device and the natural representation of the data do not always coincide, therefore, general estimation methods need to account for observations and inference in either domain (or both). The second reason is a conceptual one and it follows as a consequence of the \emph{Fourier uncertainty principle} (or Gabor limit, see the Introduction of \cite{hall2006resolution}). This principle  states that a function cannot be simultaneously localised or concentrated both in time and frequency, meaning that it has to be \emph{spread out} in at least one of those domains. Therefore, signals can be efficiently represented by freely choosing their representation domain: frequency or time indistinctly (or even parts of both domains).} 

We model Fourier pairs using probabilistic generative models \cite[Ch.~4-5]{Murphy2012}, thus being able to perform statistical inference for a broad class of signals and their spectra. This treatment blends two perspectives to spectral analysis: on one hand, the probabilistic treatment allows us to include structural information which can then be updated in the light of observations via Bayesian updates, thus accounting for model uncertainty and missing data. On the other hand, the generative perspective allows for flexible priors that can account for a large class of signals, without the stringent limitation of parametric structures. We will define a joint probabilistic model for the spectral and temporal representations of a signal; this way, whenever observations of any representation become available, all unobserved quantities can be estimated by means of Bayesian inference. \revised{By posing the spectral and temporal reconstruction problems as (only) one of Bayesian inference, the main challenge becomes to find an appropriate joint prior for both representations that ensures exact inference thus preventing the need for computationally expensive and inaccurate posterior approximations. Exploiting the fact that the representations are linked by the Fourier transform, we assume a jointly Gaussian model, since normal random variables are closed under the Fourier transform---which is a linear operator. This way, we can interpret that both representations are two components of the same object (the signal) and therefore they can be inferred using observations from either domain through closed-form (Gaussian) Bayesian inference.} Therefore, we refer to our method as \emph{Bayesian reconstruction of Fourier pairs} (BRFP). 

\revised{
We summarise the contributions of our work as follows:
\begin{itemize}
    \item A joint probabilistic model for signals and their spectra is proposed, with likelihoods that cater for noisy and missing data.
    \item Based on the above model, a mechanism for addressing spectrum and temporal reconstruction using observations from either time or frequency domains, via closed form Bayesian inference is introduced.
    \item Due to its Bayesian treatment, a natural account of uncertainty of estimates in the form of analytic error bars on both the signal and its spectrum is established.
    \item An analysis of the covariance structure of the proposed model and how it relates to classical spectral estimation is provided. In particular, for the complete, noiseless-data scenario, we show that the proposed methodology collapses to the discrete Fourier transform.
    \item Illustrative examples on synthetic data as well as a comparison to classic and modern methods in the literature on real-world datasets from astronomy and health are presented. These experiments are evaluated through appropriate performance indices. 
\end{itemize}
}

\textbf{Organisation of the paper.} Section \ref{sec:model} presents the construction of the proposed generative model in time and frequency, while Section \ref{sec:with_noise} equips the model with an observation stage dealing with missing and/or noisy data. These sections i) compare BRFP to the classical approaches, ii) analyse the resulting (posterior) covariances, and iii) provide illustrative examples. Section \ref{sec:simulations} shows the experimental validation of the proposed BRFP model against classical and modern methods in three settings using real-world data. \revised{Section \ref{sec:discussion} presents a discussion of the experimental results, while Section \ref{sec:conclusion} provides the concluding remarks and future work. Lastly, the Appendix complements the experimental section giving a brief description of the benchmarks and the 2D implementation of the proposed method.}

%!TEX root = ./BRFP.tex
\section{A Joint Model for Signals and Spectra}
\label{sec:model}

Defining generative models for variables that are strongly coupled is challenging in general. In our case, the signal and its spectrum are indeed heavily interrelated: their relationship, given by the Fourier transform, is deterministic. We will take advantage of this fact and first define the marginal distribution of the time series. Then, we will compute the corresponding joint distribution over both time and spectrum as the \emph{push-forward distribution}\footnote{The push-forward distribution is simply the distribution induced by applying a deterministic function to a random variable; in our case, the application of the Fourier transform to the time series. The name comes from the interpretation that the probability mass is being \emph{pushed forward} from the original space (time) to the transformed one (frequency).} obtained through the Fourier transform applied to time series.

\subsection{Multivariate normal prior for temporal signals}
\label{sec:model_time}
We focus on the discrete-time (or discrete-space in the case of images) setting, that is, we consider finite-length series $\x = [x_0, \dots, x_{N-1}]^\top\in\R^N$---where the subscript of $x_i\in\R$ denotes the time index and $N\in\N$. In statistical terms, we will assume signals to be realisations of a multivariate normal (MVN) distribution with  (unknown) mean $\m\in\R^N$ and covariance matrix $\Sigma\in\R^{N\times N}$, that is, 
\begin{align}
  \mathbf{x} \sim \MVN(\m,\Sigma). \label{eq:time_series} 
\end{align}

This is a probabilistic generative model for the time series $\x$, where the hyperparameters $\m$ and $\Sigma$ determine the properties of the signal such as magnitude, smoothness, and memory, to name a few. We emphasise that the assumption of normality is in fact very general, mainly because by controlling the covariance function we can account for arbitrarily-general structures, akin to Gaussian processes \cite{Rasmussen:2005:GPM:1162254}. As a matter of fact, several methods either implicitly assume Gaussianity, or their scope can be reached by the Gaussian assumption. For instance, ARMA models assume Gaussian errors and the Lomb-Scargle method~\cite{Lomb1976, Scargle1982} reconstructs the missing values by fitting a sum of sinusoids, which can be replicated with an MVN with periodic covariance function \cite[eq.~(17)]{mackay_gp}. Also, recall that the Gaussian distribution provides the largest entropy among the distributions with fixed variance supported on $[-\infty,\infty]$, therefore, the Maximum-entropy Principle supports choosing the Gaussian distribution. Lastly, the Central Limit Theorem also motivates the Gaussian assumption, since multiple sources of uncertainty added together result in a distribution that converges to a Gaussian even if the original sources followed a different distribution.

Therefore, the above argument suggests that assuming a nonparametric Gaussian model yields---by construction---more general models than a number of classical approaches. This is due to the large literature on covariance design \cite{TobarTuner2015,nips17,tobar19b}, which allows to Gaussian models to replicate well-known models in the literature. However, we acknowledge that real-world data such as the number of words in a stream of text \cite{tobar18a}, a price series in finance \cite{tobar20a} or discrete-state series can exhibit non-Gaussian features. These non-Gaussian signals can be modelled through a latent Gaussian model with an appropriate non-Gaussian likelihood that ensures the desired properties such as non-negativity, skewness or kurtosis. There is a vast literature on Gaussian models with non-Gaussian likelihoods that are used for classification \cite[Ch.~3]{Rasmussen:2005:GPM:1162254} and regression \cite{ijcnn18}, which can be motivated from neural networks \cite{rios19}, parametric non-linear functions \cite{warped04} and even Gaussian processes \cite{bayesianwarped12} or Gaussian mixtures. Therefore, as we focus on modelling the temporal dependence of signals and exploiting it in the spectral analysis setting, we argue that the Gaussian assumption is pertinent as it can be complemented with a non-Gaussian likelihood if required; however, such extensions are beyond the scope of this work. 

\subsection{The induced generative model in time and frequency}
\label{sec:model_freq}
Based on the model for (the temporal representation of) signals introduced in eq.~\eqref{eq:time_series}, we can, informally, \emph{pass this model through the Fourier transform} to calculate the distribution over the spectral representation induced by the above model. We are then interested on the push-forward measure of the Fourier spectrum of $\x$, denoted by $\X=[X_0,\ldots,X_{N-1}]$, that is,
\begin{equation}
  \X = \text{DFT}(\x), \quad   \mathbf{x} \sim \MVN(\m,\Sigma),
\end{equation}
where the expression $\text{DFT}(\cdot)$ denotes the discrete Fourier transform operator given by 
\begin{equation}
  X_k = \frac{1}{\sqrt{N}}\sum_{n=0}^{N-1} e^{-j 2\pi nk /N} x_n,
  \label{eq:dft}
\end{equation}
with $j$ denoting the imaginary unit, $k\in\{0,1,\ldots,N-1\}$ the frequency index, and  $n\in\{0,\ldots,N-1\}$ the time index. In vector form, we can express the spectrum $\X$ as~\cite{Rao2000} 
\begin{equation}
  \X = W^\top\x,
\end{equation}
where $W\in\R^{N\times N}$ is the Fourier matrix given by
\begin{equation}
  [W]_{nk}=\frac{e^{-j 2\pi nk /N}}{\sqrt{N}} = \frac{\cos(2\pi nk /N) - j\sin(2\pi nk /N)}{\sqrt{N}}\label{eq:fourierOp}.
\end{equation}
Furthermore, in order to avoid the treatment of complex-valued random variables~\cite{MandicGoh,icassp15}, we identify the real and imaginary parts of the Fourier matrix respectively by,
\begin{align}
   \Wr &= \Re W,\ \text{i.e., }[\Wr]_{nk} = \tfrac{1}{\sqrt{N}}\cos(2\pi nk /N), \label{eq:Wr}\\
   \Wi &= \Im W,\ \text{i.e., }[\Wi]_{nk} = \tfrac{-1}{\sqrt{N}}\sin(2\pi nk /N). \label{eq:Wi}
 \end{align}
This allows us to represent the spectrum of $\x$ as two real-valued vectors, corresponding to its real and imaginary parts, respectively,
 \begin{equation}
   \Xr = \Wr^\top\x,\quad \Xi = \Wi^\top\x.
   \label{eq:real_imag}
 \end{equation}

 Observe that $\Xr$ and $\Xi$ in eq.~\eqref{eq:real_imag} are simply real-valued linear transformations of the time series $\x$. Therefore, the MVN model for $\x$ in eq.~\eqref{eq:time_series} results in the vectors $\x$, $\Xr$ and $\Xi$ being jointly Gaussian as well, that is,
\begin{align}
  \left[\begin{array}{l}
    \x\\
    \Xr\\
    \Xi
  \end{array}\right] & \sim \MVN(\m_{\text{joint}}, \Sigma_{\text{joint}}),
  \label{eq:gen_mod}
\end{align}
where the mean vector and covariance matrix, respectively $\m_{\text{joint}}$ and $\Sigma_{\text{joint}}$, can be directly calculated as  
\begin{align}
  m_{\text{joint}}
  & = 
    \left[\begin{array}{c}
      \m\\
      \Wr^\top \m\\
      \Wi^\top \m
      \end{array}\right]
    = \Wa^\top \m,\label{eq:mx}\\
  \Sigma_{\text{joint}}  & = \left[ \begin{array}{ccc}
                        \Sigma & \Sigma \Wr & \Sigma \Wi \\
                        \Wr^\top \Sigma & \Wr^\top \Sigma \Wr & \Wr^\top \Sigma \Wi\\
                        \Wi^\top \Sigma & \Wi^\top \Sigma \Wr & \Wi^\top \Sigma \Wi
                        \end{array}\right]
      =
      \Wa^\top \Sigma \Wa.\label{eq:Sx}
\end{align}
We have denoted $\Wa = [\I_N\ \Wr\ \Wi]$ as the concatenated matrix comprising $\I_N$ (the identity matrix of dimension $N$, the length of the time series), and the real and imaginary Fourier matrices. 

From eq.~\eqref{eq:Sx}, it can be shown that the rank of $\Sigma_{\text{joint}}$ is that of $\Sigma$ (usually $N$) and so are its diagonal blocks corresponding to the covariance matrices of $\x$, $\Xr$ and $\Xi$.  In this sense, the above model then defines a joint prior distribution over a discrete-time series and its (2D real-valued representation) discrete-frequency spectra, where the linearity of the Fourier transform and the assumption of Gaussianity allows for closed-form expressions.

\subsection{Consistency with classical definitions}
\label{sec:consistency}

From a classical Fourier-spectrum perspective, the idea of a \emph{joint distribution} for a signal $\x$ and its spectrum $\X$ might sound counter-intuitive. This is because the Fourier transform establishes a (deterministic) bijection between the temporal and spectral representations, meaning that for a given time series $\x$, one and only one spectrum $\X$ exists. 

A direct consequence of this one-to-one relationship between a time series and it spectrum is that, if one of the representations is fully known, the conditional distributions $p(\x|\X)$ and $p(\X|\x)$ are \textit{degenerate}: they are greater than zero only for a single element (given by the Fourier transform). To show that the proposed generative model complies with this fact, let us analyse the conditional distribution of the spectrum given the time series $p(\X|\x)$---the analysis of $p(\x|\X)$ is analogous. Applying conditioning and marginalisation to eq.~\eqref{eq:gen_mod}, observe that $p(\X|\x)$ is Gaussian with mean and variance given respectively by:
\begin{equation}
  \left[\begin{array}{c}
      \Wr^\top \\
       \Wi^\top 
      \end{array}
  \right]
  \m
  +
  \left[\begin{array}{c}
      \Wr^\top \Sigma\\
       \Wi^\top \Sigma
      \end{array}
  \right]
  \Sigma^{-1}
  \left(\x-\m\right)
  =
  \left[\begin{array}{c}
      \Wr^\top\\
       \Wi^\top
      \end{array}
  \right] \x,
\end{equation}

\begin{equation}
  \left[\begin{array}{cc}
      \Wr^\top \Sigma \Wr & \Wr^\top \Sigma \Wi\\
      \Wi^\top \Sigma \Wr  &  \Wi^\top \Sigma \Wi
      \end{array}
  \right]
  -
  \left[\begin{array}{c}
      \Wr^\top \Sigma\\
       \Wi^\top \Sigma
      \end{array}
  \right]
  \Sigma^{-1}
  [\Sigma\Wr, \Sigma\Wi]
  %\left[\begin{array}{c}
  %    \Wr \Sigma\\
  %     \Wi \Sigma
  %    \end{array}
  %\right]
  =
  0.
\end{equation}
Therefore, the posterior distribution of the spectrum $\X$ given the entire time series $\x$ concentrates on the (deterministic) DFT with zero variance. 

The exclusive advantage of the proposed generative model arises in scenarios where the time series, or the spectrum, is not fully known but only available through incomplete and/or corrupted observations---see Section \ref{sec:with_noise}. Before that, Sections \ref{sec:covariance_analysis} and \ref{sec:sampling_prior} are dedicated to analysing the resulting covariances and how to sample from the proposed model respectively.

\subsection{Analysis of the induced covariances}
\label{sec:covariance_analysis}

As the joint temporal/spectral generative model above is Gaussian, all of its structure is given by its covariances since the mean only acts as a vertical shift in time. The analysis of these covariances is key for sampling and computing posterior distributions. 

Let us define the (marginal) covariances of the real and imaginary parts of the spectrum as $\Kr$ and $\Ki$ respectively. Then, notice from eq.~\eqref{eq:Sx}, that these covariances are quadratic forms of the Fourier operators given (coordinate-wise) by\footnote{We use the following notations for the $(p,q)^{\text{th}}$ element of a matrix $A$ indistinctly: the explicit notation $A(p,q)$, and the compact notation $[A]_{p,q}$. Accordingly, we denote the $q^{\text{th}}$ column of $A$ by $[A]_{:q}$, the $q^{\text{th}}$ row of $A$ by $[A]_{q:}$, and the $p^{\text{th}}$ element of a vector $V$ simply as $V(p)$.}  
\begin{alignat}{3}
	\Kr(k,k') &= [\Wr^\top]_{k:}\Sigma [\Wr]_{:k'} &&= [\Wr]^\top_{:k}\Sigma [\Wr]_{:k'}\label{eq:Kr}\\
	\Ki(k,k') &= [\Wi^\top]_{k:}\Sigma [\Wi]_{:k'} &&= [\Wi]^\top_{:k}\Sigma [\Wi]_{:k'}\label{eq:Ki},
\end{alignat}
where $k,k'\in\{0,1,\ldots,N-1\}$ are two arbitrary frequency indices according to the definition of the DFT in eq.~\eqref{eq:dft}. 

The first key fact about the covariances $\Kr$ and $\Ki$ is that they are not (necessarily) stationary, i.e., it is not guaranteed that $\Kr(k,k') = \Kr(k-k')$. Therefore, the power %\cyan{\bf[WHAT IS MEANT BY POWER HERE?]} 
of the random variable $\Xr(k)$ does depend on the frequency index $k$, and the same holds for the imaginary part $\Xi(k)$. This can be understood from the Hermitian forms in eqs.~\eqref{eq:Kr}-\eqref{eq:Ki}, which are equivalent to weighted inner products between the Fourier operators (at frequencies $k$ and $k'$) with weights given by the  matrix $\Sigma$. This inner product is maximised (wrt to $k$ and $k'$) when the Fourier operator is proportional, or co-linear, to the eigenvectors of $\Sigma$. Therefore, $\Kr(k,k)$ depends on the value of $k$, which contradicts stationarity. 

Additionally, since  $\Wr$ is even and $\Wi$ is odd---see eqs.~\eqref{eq:Wr} and \eqref{eq:Wi}---the covariances $\Kr$ and $\Ki$ are also  even and odd respectively on each component:\footnote{For notational consistency, we clarify that negative frequency indices follow from choosing negative frequencies in the definition of the Fourier matrices in eqs.~\eqref{eq:Wr}-\eqref{eq:Wi}.} 
\begin{align}
	\Kr(k,k') &= \Kr(-k,k')= \Kr(k,-k')= \Kr(-k,-k'),\\
	\Ki(k,k') &= -\Ki(-k,k')= -\Ki(k,-k')= \Ki(-k,-k').
\end{align}
These relationships between the covariances translate directly to the spectrum: we can easily verify that
\begin{align}
  p(\Xr(-k)|\Xr(k)) &= \Normal{\Xr(k),0} = \delta_{\Xr(k)},\\
  p(\Xi(-k)|\Xi(k)) &= \Normal{-\Xi(k),0}  = \delta_{-\Xi(k)},
\end{align}
which means that $\Xr$ is even and $\Xi$ is odd with probability 1 (i.e., \emph{almost surely}). This implies that the negative (cf.~positive) part of the spectrum uniquely determines the positive (cf.~negative) one.   

We now analyse the (cross-)covariances between the temporal and spectral representations. Denoted by $\Ktr = \Sigma \Wr$ (cf.~$\Kti=\Sigma \Wi$), the covariance between the time series $\x$ and the real $\Xr$ (cf.~imaginary $\Xi$) part of the spectrum---see eq.~\eqref{eq:Sx}---are given coordinate-wise by %\cyan{\bf[in the next two expressions we could add $2\pi$ under $\cos$ and $\sin$.]}:
\begin{align}
\Ktr(n,k) &=  [\Sigma]_{n:} [\Wr]_{:k} = \tfrac{1}{\sqrt{N}}\sum_{p=1}^N[\Sigma]_{np} \cos(2\pi pk/N),\\
\Kti(n,k) &=  [\Sigma]_{n:} [\Wi]_{:k} = \tfrac{1}{\sqrt{N}}\sum_{p=1}^N[\Sigma]_{np} \sin(2\pi pk/N).
 \end{align}
Observe that, besides the even/odd property of $\Ktr$ and $\Kti$ given by the sum of sinusoids, the inner product structure above suggests that the covariances  are large whenever the $n^\text{th}$ column (or row, due to symmetry) of the covariance function oscillates at the $k^\text{th}$ frequency. This is rather intuitive: a time series is highly correlated with its spectrum at a given frequency $k$ if the signal itself has energy at that frequency.

\subsection{Sampling from the proposed model}
\label{sec:sampling_prior}

As it is customary in the generative models literature, we now focus on sampling from the proposed model, that is, on the random generation of pairs of (synthetic) signals and their corresponding spectrum. We clarify that in this section, \emph{sampling} is regarded in the statistical sense of drawing realisations from a distribution (or statistical model), and not in the common signal processing sense related to data acquisition. The main challenge related to sampling from the proposed model stems from the strong (or, in fact, deterministic) relationship between both representations, which makes joint sampling difficult. 

Specifically, sampling from $[\x,\Xr,\Xi]$ directly should be avoided, since the deterministic coupling among such quantities can lead to numerical instabilities arising from the noninvertibility of $\Sigma_{\text{joint}}$ in eq.~\eqref{eq:Sx}, which is not full rank as explained in Sec.~\ref{sec:consistency}. Therefore, we sample from $p(\x,\X)$ in a hierarchical manner, that is, by first sampling $\x\sim p(\x)$, in eq.~\eqref{eq:time_series}, and then $\X|\x\sim p(\X|\x)$. Luckily, in our case this last distribution is a Dirac of the form $p(\X|\x) = \delta_{W^\top \x}(\X)$, as a consequence, sampling $\X|\x$ only amounts to calculating $\X = W^\top \x$. 

\begin{figure}[t]
    \centering
    \includegraphics[width = 0.7\linewidth]{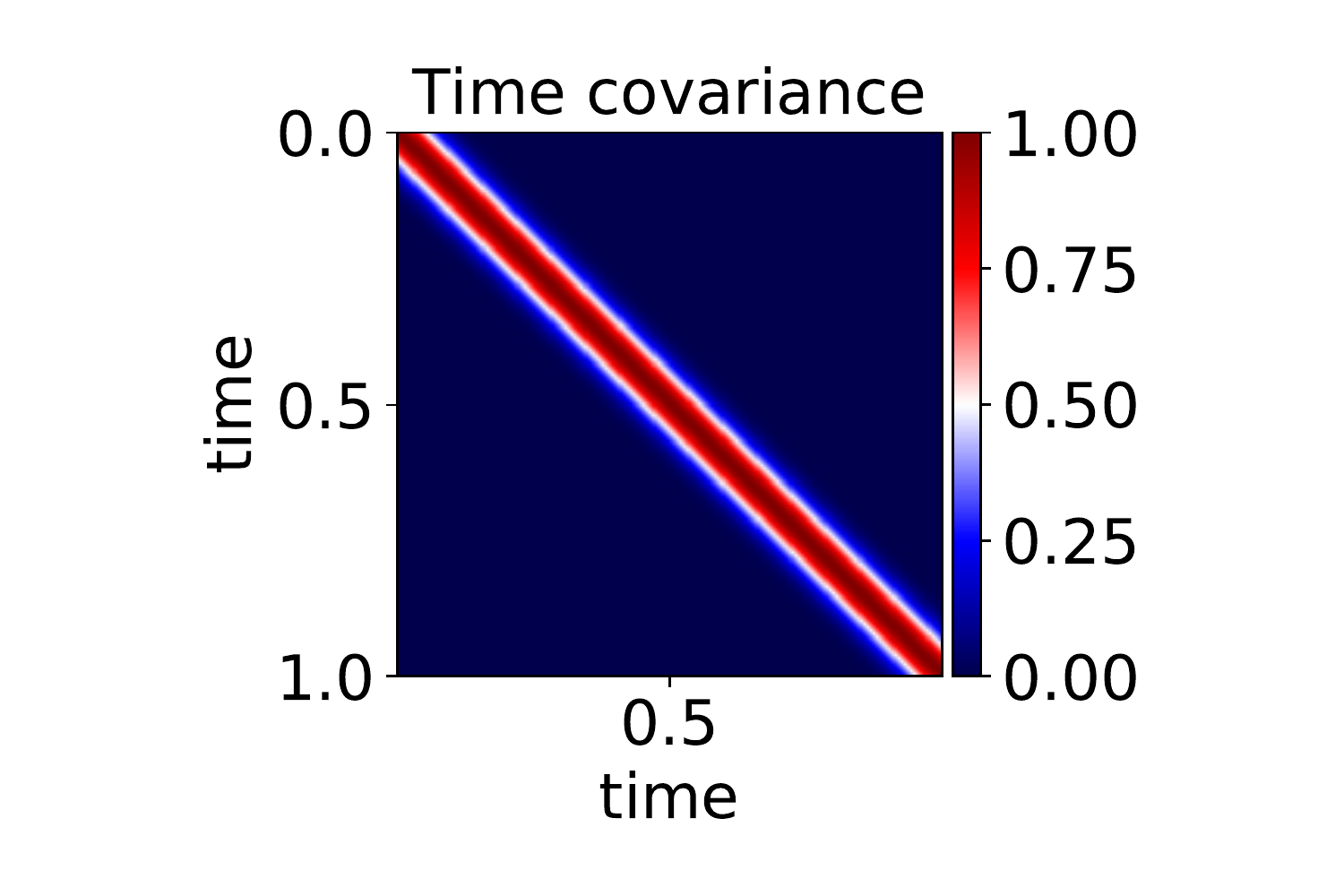}
    \caption{Temporal covariance matrix $\Sigma$ given by an SE kernel with hyperparameters $\alpha = 0.001\times 512^2$, $\sigma = 1$.}
    \label{fig:time_covariance}
\end{figure}

\begin{figure}[t]
    \centering
    \begin{subfigure}[t]{\linewidth}
        \includegraphics[width=\linewidth, trim={0 0 0 30pt}, clip]{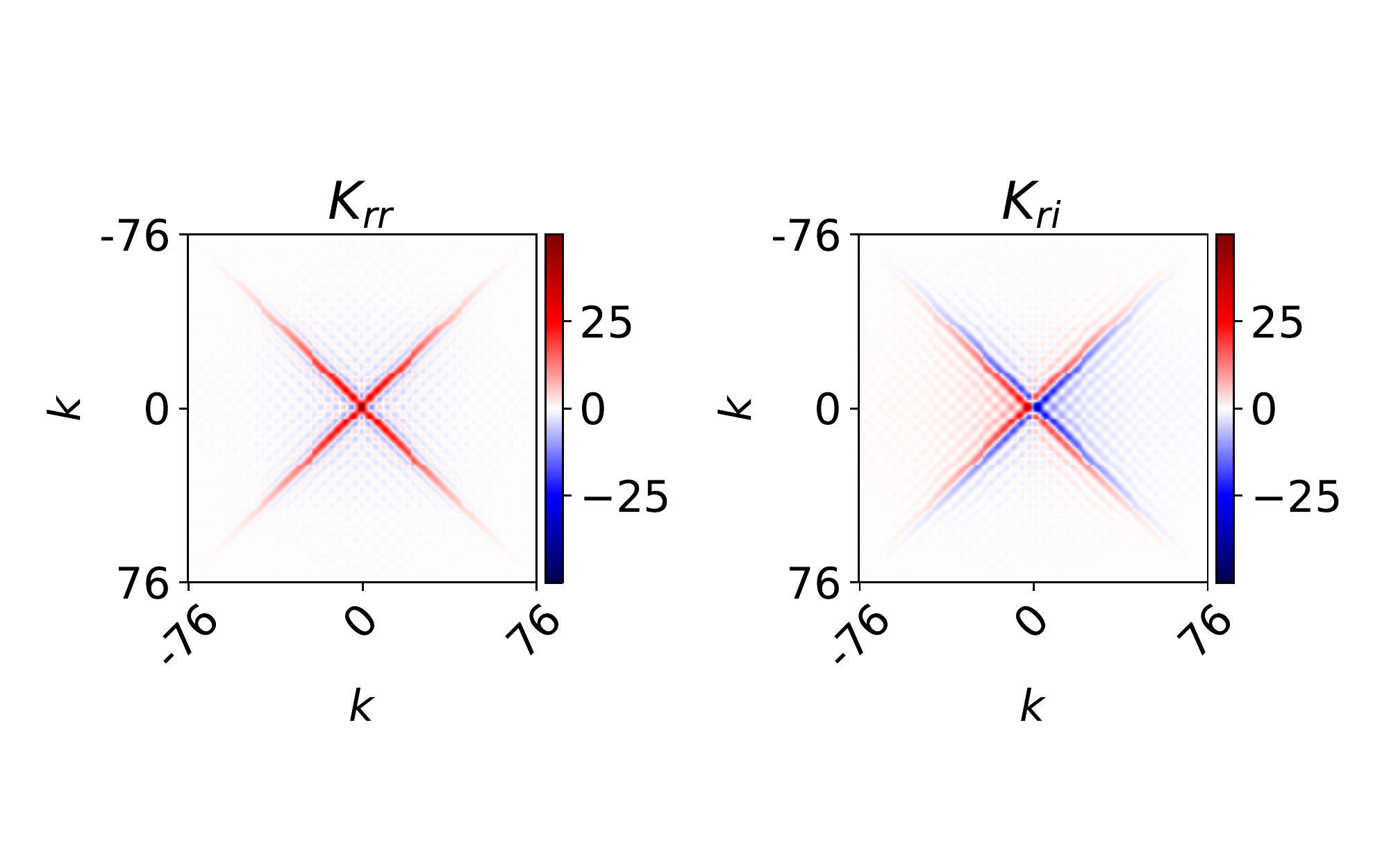}
    \end{subfigure}\vspace{-35pt}
    \begin{subfigure}[t]{\linewidth}
        \includegraphics[width=\linewidth, trim={0 40pt 0 50pt}, clip]{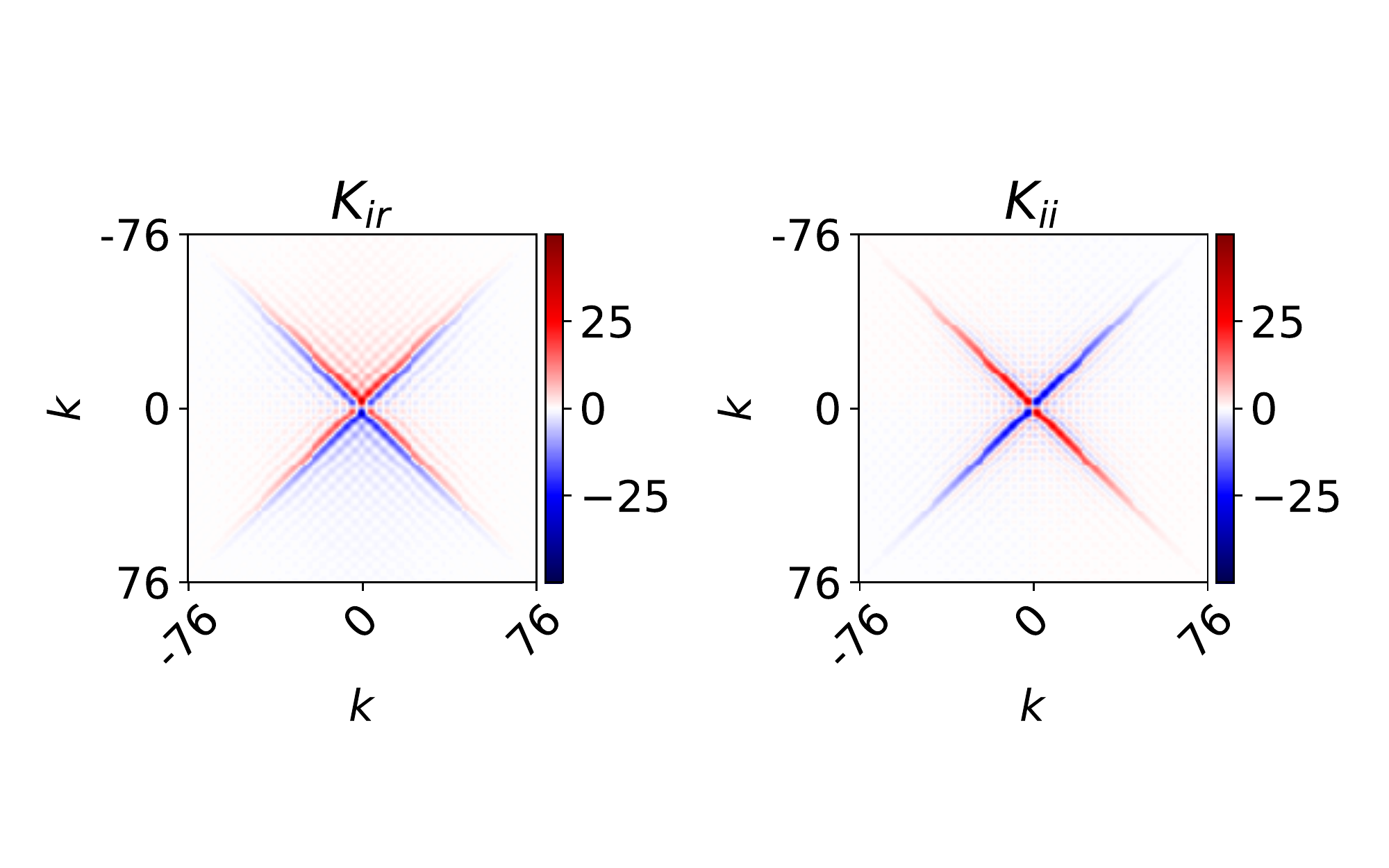}
    \end{subfigure}
    \caption{Spectrum covariance matrices induced by choosing an SE temporal covariance. Since the energy of the spectrum using an SE kernel is concentrated in the origin, these matrices have been zoomed in for frequency indices in the range [-76,76].}
    \label{fig:spectrum_covariance}
\end{figure}

\begin{figure}[t]
    \centering
    \begin{subfigure}[t]{0.5\linewidth}
        \includegraphics[width=\linewidth]{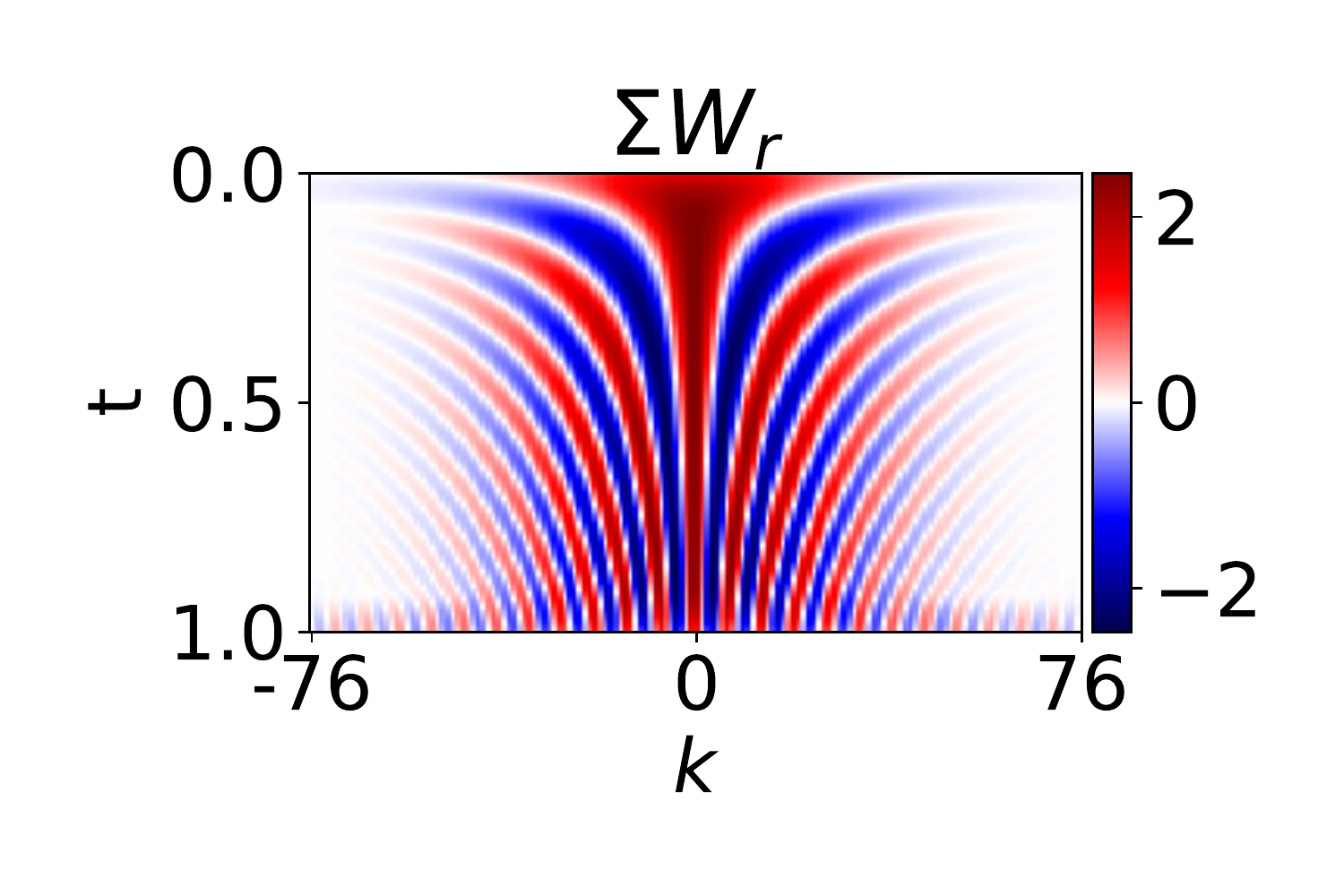}
    \end{subfigure}%
    \begin{subfigure}[t]{0.5\linewidth}
        \includegraphics[width=\linewidth]{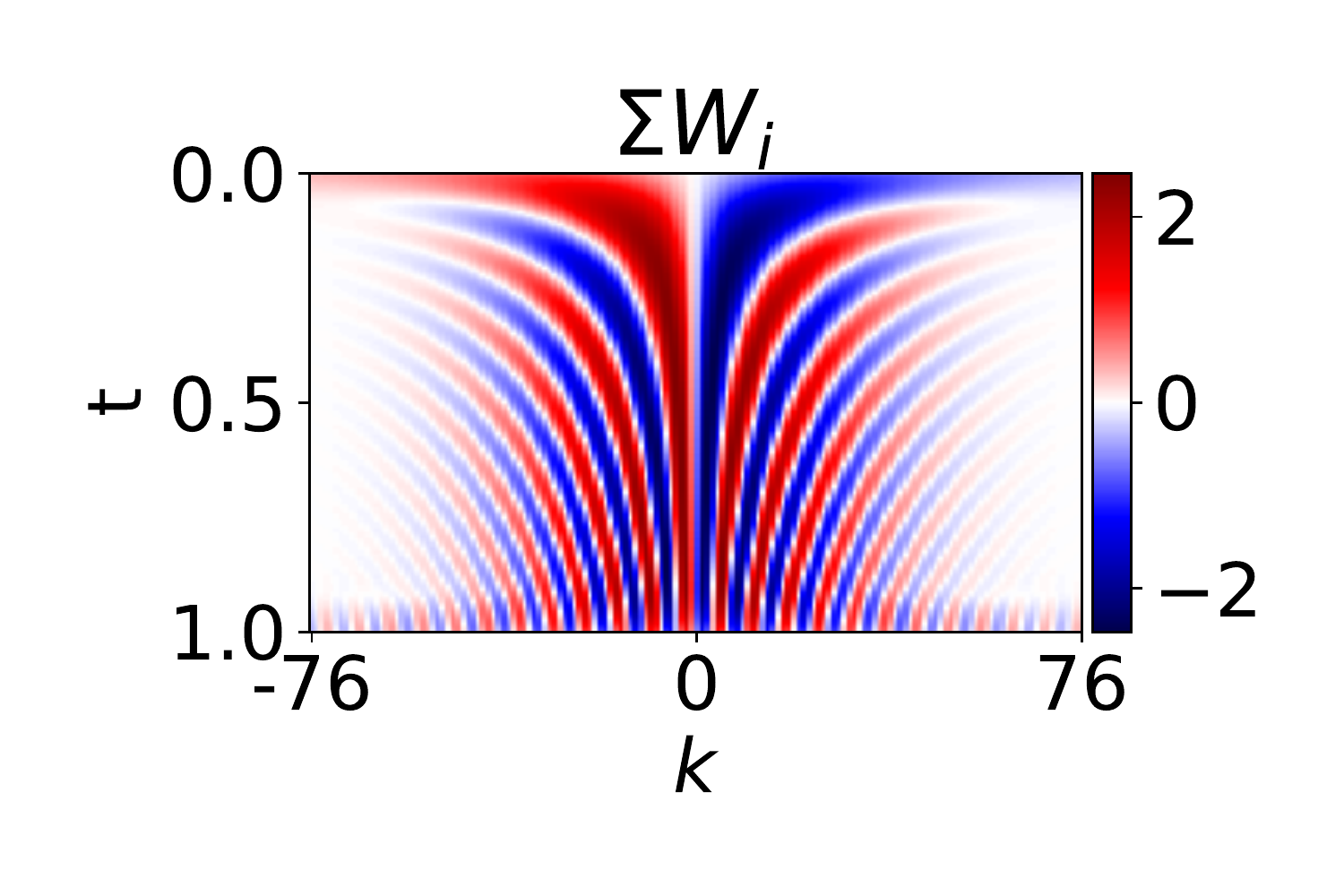}
    \end{subfigure}
    \caption{Covariances between the time series and its spectrum choosing an SE temporal covariance: real (left) and imaginary (right) parts. The frequency range has been zoomed in as well.}
    \label{fig:time_spectrum_covariance}
\end{figure}

We illustrate the above sampling rationale with a Gaussian example. To define the covariances, we chose  a time grid given by $512$ evenly-sampled values in the interval $[0, 1]$ and parametrised the temporal covariance $\Sigma$ by the \emph{square exponential} (SE) kernel \cite{Rasmussen:2005:GPM:1162254}:
\begin{equation}
  \label{eq:SE_kernel}
	[\Sigma]_{t,t'} = \sigma^2\exp\left(-\alpha(t-t')^2\right),
\end{equation}
where the hyperparameters $\sigma^2$ represent the covariance magnitude and $\alpha>0$ the \emph{rate}. Fig.~\ref{fig:time_covariance} shows the SE kernel for the chosen time grid and hyperparameters; notice that this temporal covariance $\Sigma$ is stationary. 
Then, following eq.~\eqref{eq:Sx}, we computed all joint covariances of the time series and the real and imaginary parts of the spectrum. Fig.~\ref{fig:spectrum_covariance} shows the covariances of the real and imaginary parts,  $\Kr$ and $\Ki$, which obey the symmetric properties described in Section \ref{sec:covariance_analysis}. Lastly, Fig.~\ref{fig:time_spectrum_covariance} shows the joint covariances between the temporal and spectral representations.

\begin{figure}[t]
    \centering
    \includegraphics[width = \linewidth]{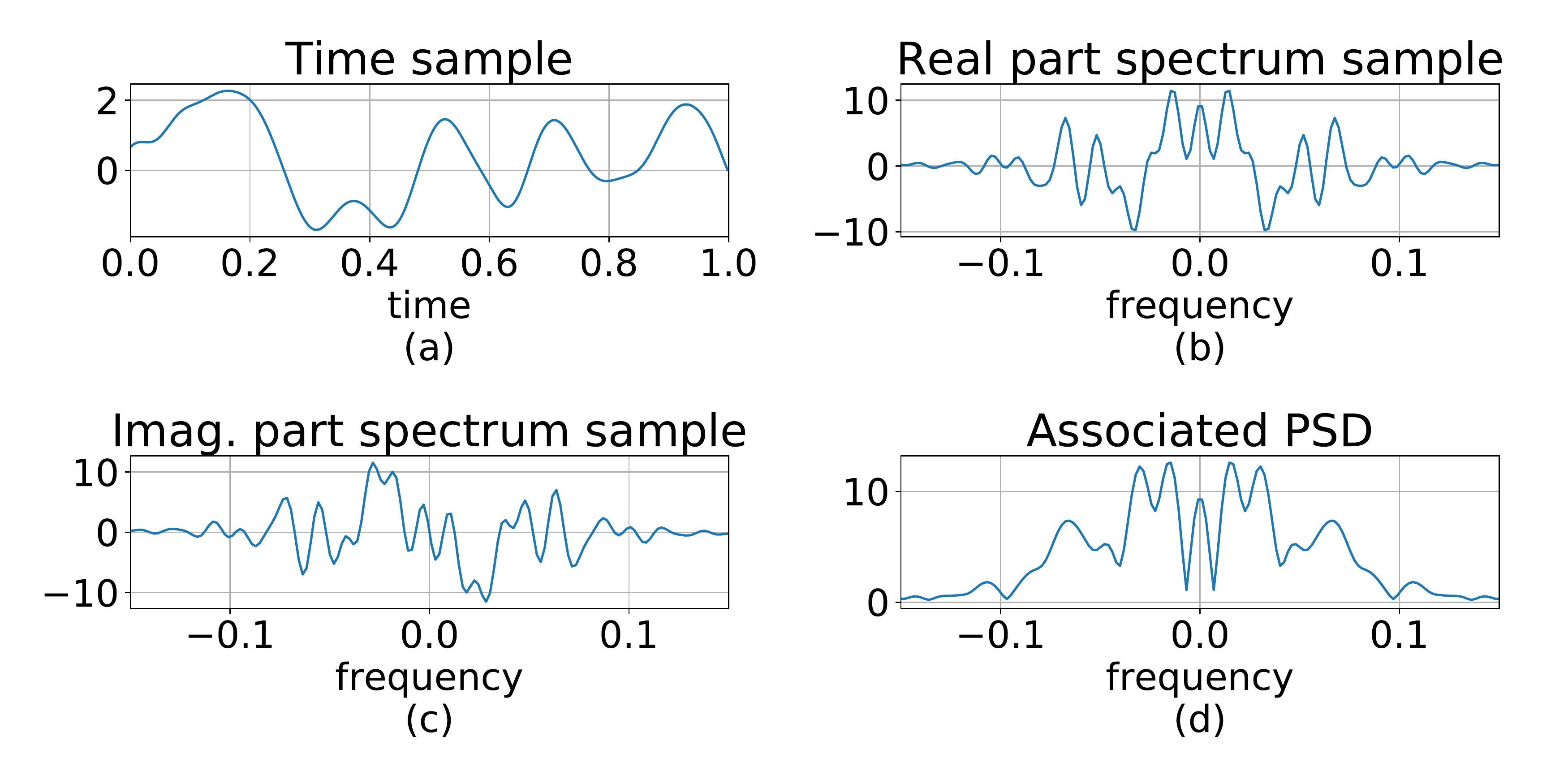}
    \caption{A sample from the proposed generative model. (a) Time series, (b) real part of spectrum, (c) imaginary part of the spectrum and (d) power spectral density.}
    \label{fig:signals}
\end{figure}

With the defined covariances we sample hierarchically as described above. Fig.~\ref{fig:signals} shows the time and frequency samples (real and imaginary parts), together with the power spectral density (PSD) given by the square absolute value of the spectral representation. Notice that the real part of the spectrum is even whereas the imaginary part is odd as discussed in Section \ref{sec:covariance_analysis}. 

%!TEX root = ./BRFP.tex

\section{Reconstruction under missing/corrupted observations}
\label{sec:with_noise}
In this section, we extend the proposed model to deal with real-world scenarios, where observations may be missing and/or contaminated with noise. To this end, we first equip the proposed model with a likelihood that allows for such cases, to then focusing on training (i.e., finding the model parameters) and inference (i.e., removing noise and imputing missing values). We conclude this section with an illustrative example.

\subsection{Observation model}
\label{sec:observation_model}

In order to cater for time series that cannot be observed directly, we consider the $\Mt$-dimensional observation vector $\y\in\R^{\Mt}$ as a noise-corrupted linear transformation of the time series $\x\in\R^N$ given by
\begin{align}
  \y & = \Ht^\top \x + \epsilon\label{eq:y},  
\end{align}
where $\epsilon$ is a Gaussian white-noise vector with covariance $\st^2 \I_{\Mt}$, and  $\Ht$ is an observation matrix that encodes how the values of the temporal representation $\x$ are obtained through values of the temporal observation $\y$. Although the proposed methodology admits an arbitrary observation matrix $\Ht\in\R^{N\times\Mt}$, we focus on the missing data case, where only one value of the time series is observed through each entry of $\y$, and some values of $\x$ are not observed whatsoever. That is, 
\begin{align}
\label{eq:Hx}
[\Ht]_{p,q} & = \begin{cases}
1 & \text{If } \x(p) \text{ is observed through } \y(q),\\
0 & \text{if not},
\end{cases}
\end{align}
where, necessarily, $\Mt\leq N$. 

In a similar way, we can define an observation model for the frequency domain through an observation matrix $\Hf\in \R^{N \times \Mf}$ relating the full (but latent) spectrum $\Xr, \Xi \in \R^{N}$ and a subset of $\Mf$ noisy spectrum observations given by $\Yr,\Yi\in \R^{\Mf}$ according to
\begin{align}
  \Yr & = \Hf^\top \Xr + \er\label{eq:yr},\\
  \Yi & = \Hf^\top \Xi + \ei\label{eq:yi},
\end{align}
where $\er$ and $\ei$ are Gaussian white-noise vectors of variance $\sf^2\I_{\Mf}$. We choose these noise sources to be independent, however, they can be modelled as  correlated if needed. Also, akin to $\Ht$ in eq.~\eqref{eq:Hx}, the spectral observation matrix $\Hf$ is defined, element-wise, as $[\Hf]_{p,q}  = 1$ if and only if the $p$-th entry of the spectrum is observed through the $q$-th observation, and $[\Hf]_{p,q}=0$ otherwise. We emphasise that we do not impose any relationship between $\Hf$ and $\Ht$, since they are given by temporal/spectral acquisition devices that are specific to each setting. 

\begin{figure}[t]
    \centering
    \includegraphics[width = 0.8\linewidth]{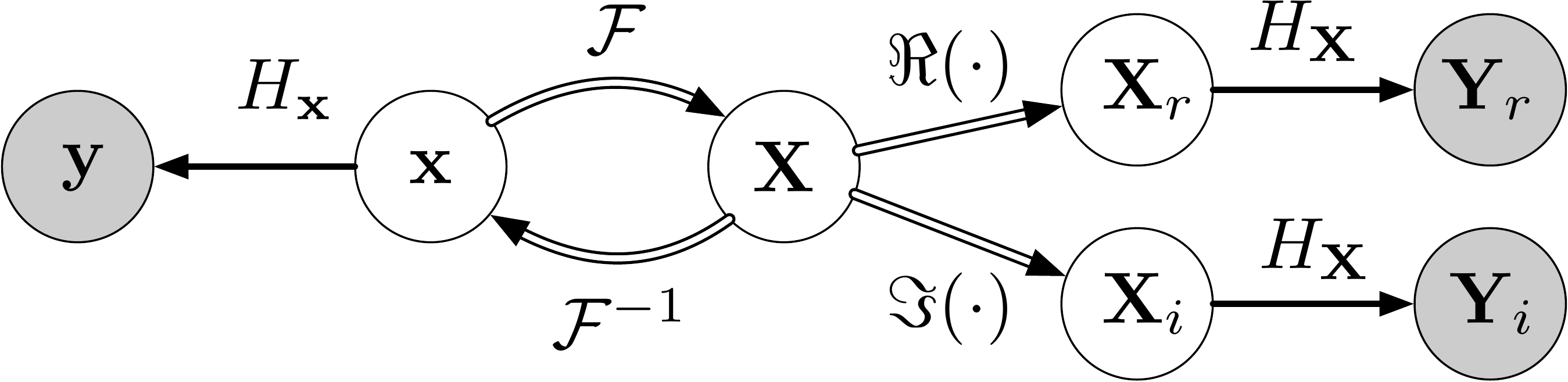}
    \caption{Graphical model representation of the joint generative model for time series, spectrum and observations.}
    \label{fig:GM}
\end{figure}

Fig.~\ref{fig:GM} shows a graphical representation of the proposed model comprising all latent variables (white nodes) in both the time and frequency domains, together with their observables (grey nodes); deterministic and stochastic dependencies are denoted by double and single lines respectively. The corresponding transformations have been denoted along each arc.

\subsection{Training and posterior computation}
With the joint prior over time and frequency defined in Secs.~\ref{sec:model_time} and \ref{sec:model_freq}, and the likelihood (observation model) in Sec.~\ref{sec:observation_model}, we  focus on training and inference using the proposed model. In our setting, this respectively refers to (i) finding the parameters of the mean and covariance functions, or the \emph{model's hyperparameters}, and (ii) computing the posterior distribution over both the time series and its spectrum, based on the set of partial, noise-corrupted observations in both domains.

Towards a simpler notation to address training and inference, let us refer to $\y,\Yr,\Yi$ as \textit{observations} and to $\x,\Xr,\Xi$ as \textit{latent variables}. In a more compact notation, the observations can be expressed as 
\begin{align}
  \label{eq:compact_obs}
  \underbrace{\left[\begin{array}{c}
      \y \\ \Yr \\ \Yi
      \end{array}
  \right]}_{\Ya}
  &=
  \underbrace{\left[\begin{array}{ccc}
      \Ht^\top & 0 & 0 \\
      0 & \Hf^\top & 0  \\
      0 & 0 & \Hf^\top
      \end{array}
  \right]}_{\Ha^\top}
  \underbrace{\left[\begin{array}{c}
      \I \\ \Wr^\top \\ \Wi^\top
      \end{array}
  \right]}_{\Wa^\top}
   \x
   +
   \underbrace{\left[\begin{array}{c}
       \epsilon \\ \er \\ \ei
       \end{array}
   \right]}_{\epsilon}\nonumber\\
   \Ya&=\Ha\Wa\x + \epsilon,
\end{align}
where we refer to  $\Ya = [\y^\top,\Yr^\top,\Yi^\top]^\top$ as the \textit{augmented observation}.

In order to train the proposed model, we first calculate the mean and covariance of the observations, denoted respectively by $\m_{\obs}$ and $\Sigma_{\obs}$, given by\footnote{We adopt the following notation: 
$\E\left[\x\right]$ is the expected value of $\x$,
$\V\left[\x\right]$ is the covariance  of $\x$, and $\V\left[\x,\x'\right]$ is the covariance  between $\x$ and $\x'$.} 
\begin{align}
  \m_{\obs} &= \E\left[ \Ha^\top\Wa^\top\x   + \epsilon\right] = \Ha^\top\Wa^\top \m \label{eq:mean_obs}\\
  \Sigma_{\obs} &=
  \V \left[ \Ha^\top\Wa^\top\x   + \epsilon\right]
  =
   \Ha^\top\Wa^\top \Sigma \Wa\Ha + \Lambda_{\text{obs}},   \label{eq:var_obs}
\end{align}
where $\Lambda_{\text{obs}}$ is the concatenated observation noise variance given by 
\begin{equation}
   \Lambda_{\text{obs}} = \left[\begin{array}{ccc}
      \I_{\Mt}\st^2 & 0 & 0 \\
      0 & \I_{\Mf}\sf^2 & 0  \\
      0 & 0 & \I_{\Mf}\sf^2
      \end{array}
  \right],\label{eq:obs_moise_var}
 \end{equation} 
and recall that $\I_{M}$ is the $M$-by-$M$ identity matrix.
\begin{remark}
The diagonal structure of the observation noise covariance in eq.~\eqref{eq:obs_moise_var} is given by the assumption of independence of $\epsilon,\er,\ei$. However, should these noise sources be correlated, their covariances would appear in the off-diagonal blocks of $\Lambda_{\text{obs}}$.
\end{remark}

With these expressions relating $\m$ and $\Sigma$ to the statistics of the observations, training can be achieved by means of maximum likelihood, that is, by maximising the likelihood function given by 
\begin{equation}
  l(\m,\Sigma) =  p(\Ya|\m_{\obs},\Sigma_{\obs}) = \cN\left(\Ya;\m_{\obs},\Sigma_{\obs}\right),
\end{equation}
for which there is a large body of specialised derivative-based and derivative-free methods in the open literature \cite{Murphy2012}. 

We now turn to the computation of the posterior. Let us first observe that the mean and covariance of the \emph{latent quantities} are similar to those of the \emph{observed quantities} in eqs.~\eqref{eq:mean_obs}-\eqref{eq:var_obs} except for the noise variance, that is,
\begin{align}
  \m_{\lat} &= \E\left[ \Ha^\top\Wa^\top \x \right] = \Ha^\top\Wa^\top \m,\\
  \Sigma_{\lat} &=
  \V \left[\Ha^\top\Wa^\top\x \right]
  =
   \Ha^\top\Wa^\top \Sigma \Wa \Ha.   
\end{align}

Then, in order to compute the posterior distribution over the missing values of the signals (both in time and frequency), we only require the  cross-covariance between the latent variables and the observations, denoted $\Sigma_{\lat,\obs}$, given by
{\small 
\begin{align}
  \Sigma_{\lat,\obs}&=
  \V \left[\Ha^\top\Wa^\top\x + \epsilon,\Wa^\top\x\right]
  =
   \Ha^\top\Wa^\top \Sigma \Wa,
\end{align}
which yields the posterior 
\begin{equation}
  \label{eq:full_posterior}
  \MVN\left(\m_{\lat} + \Sigma_{\lat,\obs} \Sigma_\obs^{-1}(\Ya-\m_{\obs}), \Sigma_{\lat} -  \Sigma_{\lat,\obs}\Sigma_\obs^{-1}\Sigma_{\lat,\obs}^\top\right).
\end{equation}
}
In particular, we are interested in the posterior of the spectrum given observations of the time series---the reciprocal is analogous. This posterior is obtained by integrating out the temporal latent variables from eq.~\eqref{eq:full_posterior}, and therefore it is Gaussian with mean and variance (we assume zero prior mean,  $\m=0$, for notational simplicity) respectively given by
{\small  
\begin{align}
    \m_{\X|\y} & =
    \left[\begin{array}{c}
        \Wr^\top \\ \Wi^\top
        \end{array}
    \right]\Sigma \Ht(\Ht^\top\Sigma \Ht + \I_{\Mt}\st^2)^{-1}\y, \label{eq:mean_post_spectral}\\
    \Sigma_{\X|\y}&=
    \left[\begin{array}{c}
        \Wr^\top \\ \Wi^\top
        \end{array}
    \right]\left(\Sigma  - \Sigma \Ht(\Ht^\top\Sigma \Ht + \I_{\Mt}\st^2)^{-1} \Ht^\top\Sigma \right)
    \left[\begin{array}{c}
        \Wr \\ \Wi
        \end{array}
    \right]
\label{eq:var_post_spectral}.
\end{align}
}

Observe that, besides the Fourier matrices $\Wr, \Wi$, the above equations correspond to the linear estimation of the latent variable $\x$ given the observation $\y$. This suggests the following result. 
\begin{proposition}
Under the proposed model, calculating the posterior distribution of the spectral representation conditional to partial and noise-corrupted temporal observations is equivalent to calculating the posterior distribution of the latent (temporal) signal, and then applying the discrete Fourier transform.
\end{proposition}
 
\revised{The above expressions for training and inference of  BRFP allows us to analyse its computational complexity. First, the cost of computing the Gaussian likelihood (and therefore the training cost) is of order $\mathcal{O}(N^3)$, where $N$ is the number of observations; this is due  to the inverse covariance matrix present in the likelihood. Though this might be large depending on the amount of observations, the fact that the cost only depends on the amount of data is favourable against parametric models, since these might have a cost of order $\mathcal{O}(NP)$, where $P$ is the number of parameters. The absence of the amount of parameters in the cost of training BRFP reveals that the model becomes more complicated as the amount of observations increase, which validates its flexibility. Regarding the operation cost, that is, computing predictions once the model is trained, eqs.~\eqref{eq:mean_post_spectral}-\eqref{eq:var_post_spectral} reveal that the prediction cost is linear on the learnt statistics and therefore it is computationally efficient both for computing the mean and variances.}

We can thus rely on eqs.~\eqref{eq:mean_post_spectral}-\eqref{eq:var_post_spectral} to analyse how the (spectrum) posterior behaves in the following particular cases. \\

\textbf{1) Noiseless complete data} ($\Ht=\I$, $\st^2=0$). The posterior spectral mean reverts to the DFT of the observations and the variance vanishes. This is a consequence of the observations being \emph{true and complete}.\\

\textbf{2) Noisy complete data} ($\Ht=\I$, $\st^2>0$). The posterior mean in eq.~\eqref{eq:mean_post_spectral} turns into a regularised linear estimator, meaning that the posterior is a smoothed version of the DFT of the observations, where the assumption that the observations are corrupted by Gaussian noise results in a variance-based regularisation.\\

\textbf{3) Large observation noise, regardless of the available data} ($\st^2\gg1$). To analyse the behaviour of the posterior spectrum in this case, let us apply the Woodbury matrix inversion lemma to eq.~\eqref{eq:var_post_spectral} to give 
\begin{align}
    \Sigma_{\X|\y}
    &=
    \left[\begin{array}{c}
        \Wr^\top \\ \Wi^\top
        \end{array}
    \right]
    \left(\Sigma ^{-1} + \frac{1}{\st^2}\Ht\Ht^\top\right)^{-1}
    \left[\begin{array}{c}
        \Wr \\ \Wi
        \end{array}
    \right]\label{eq:var_post_spectral2}.
\end{align}
Observe that when $\st^2\gg1$, eq.~\eqref{eq:mean_post_spectral} implies that the regulariser becomes large too, and the observations $\y$ are not taken into account. Additionally, from eq.~\eqref{eq:var_post_spectral2} we verify that the term with the observation matrices $\Ht$ vanishes, therefore, in the case of  large observation noise, the posterior of the spectral representation reverts to its own prior distribution: this confirms that a large amount of noise deems the observations useless.

\subsection{Two illustrative examples}

We now illustrate the proposed \emph{Bayesian reconstruction of Fourier pairs} (BRFP) method in terms of training and inference in two scenarios using synthetic data. In the first case we recovered both time and frequency signals using partial and noise-corrupted observations from both domains. The second experiment focuses on detecting the fundamental frequencies (i.e., periodicities) of a sum-of-cosines from a limited number of samples where we compare the proposed BRFP against Lomb-Scargle \cite{Lomb1976,Scargle1982}, which is specially suited to detect line spectra (i.e., sum of sinusoids). 

\subsubsection{Joint time and frequency reconstruction}

We implemented the Gaussian generative model in eq.~\eqref{eq:time_series} for time series with zero mean ($\m=0$) and covariance $\Sigma$ given by the SE kernel as shown in eq.~\eqref{eq:SE_kernel} with parameters $\alpha = 0.001$ and $\sigma = 1$. We chose a time grid given by an uniform sampling interval $[0,1]$ with size $N=512$; we then sampled a time series, $\x$, and computed its spectrum via the DFT, $\X$. We generated the partial, noise-corrupted, observations $\Ya = [\y^\top,\Yr^\top,\Yi^\top]^\top$ by using only 2\% of randomly selected time and spectrum observations with white Gaussian noise of variances $\st^2=\sf^2=0.2$. The aim of this experiment was to implement the proposed BRFP to recover (or \emph{reconstruct}) the latent variables $\x$ and $\X$ from the augmented observation $\Ya$ as defined in eq.~\eqref{eq:compact_obs}.

\begin{figure}[t]
    \centering
    \includegraphics[width = 1\linewidth]{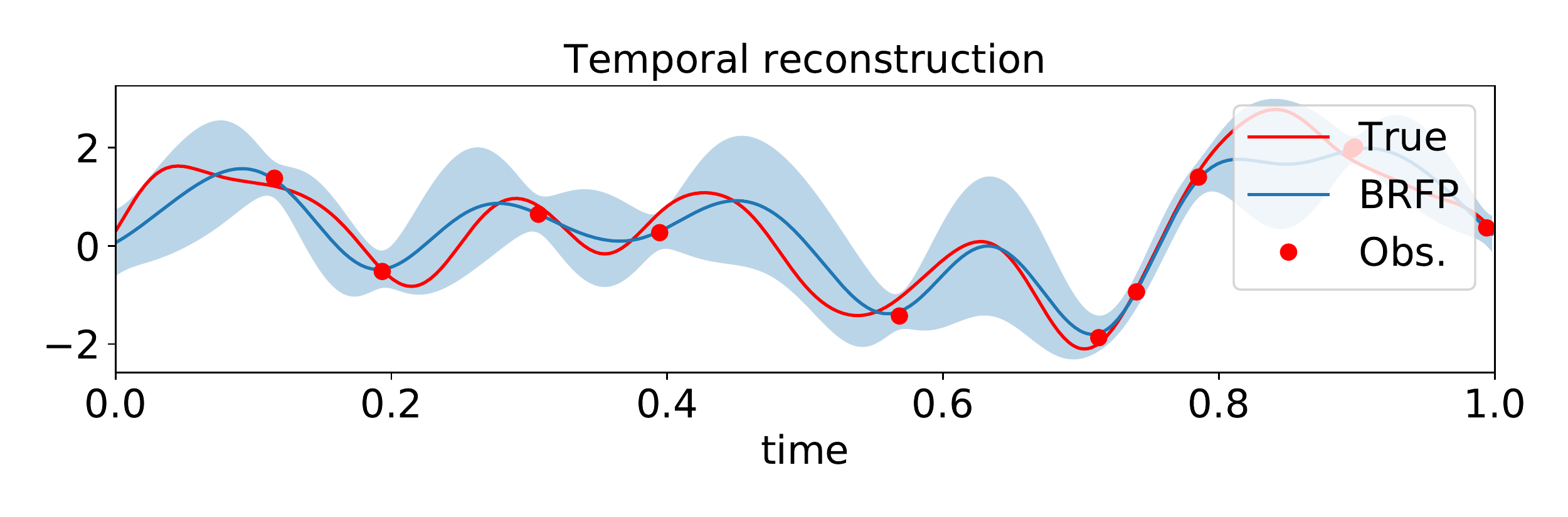}
    \caption{BRFP's temporal reconstruction of a synthetic time series using 2\% of noise-corrupted time and frequency observations. The posterior mean is in solid blue and the 95\% confidence band is in light blue.}
    \label{fig:time}
\end{figure}

\begin{figure}[t]
    \centering
    \begin{subfigure}[t]{\linewidth}
        \includegraphics[width=\linewidth]{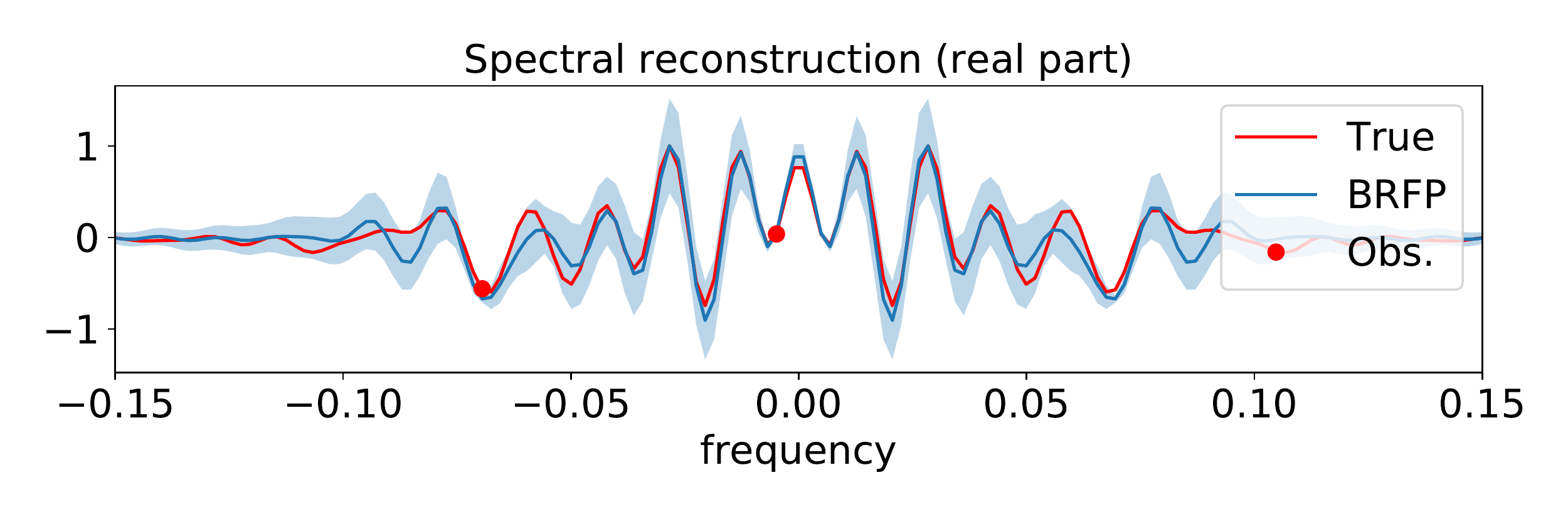}
    \end{subfigure}
    \begin{subfigure}[t]{\linewidth}
        \includegraphics[width=\linewidth]{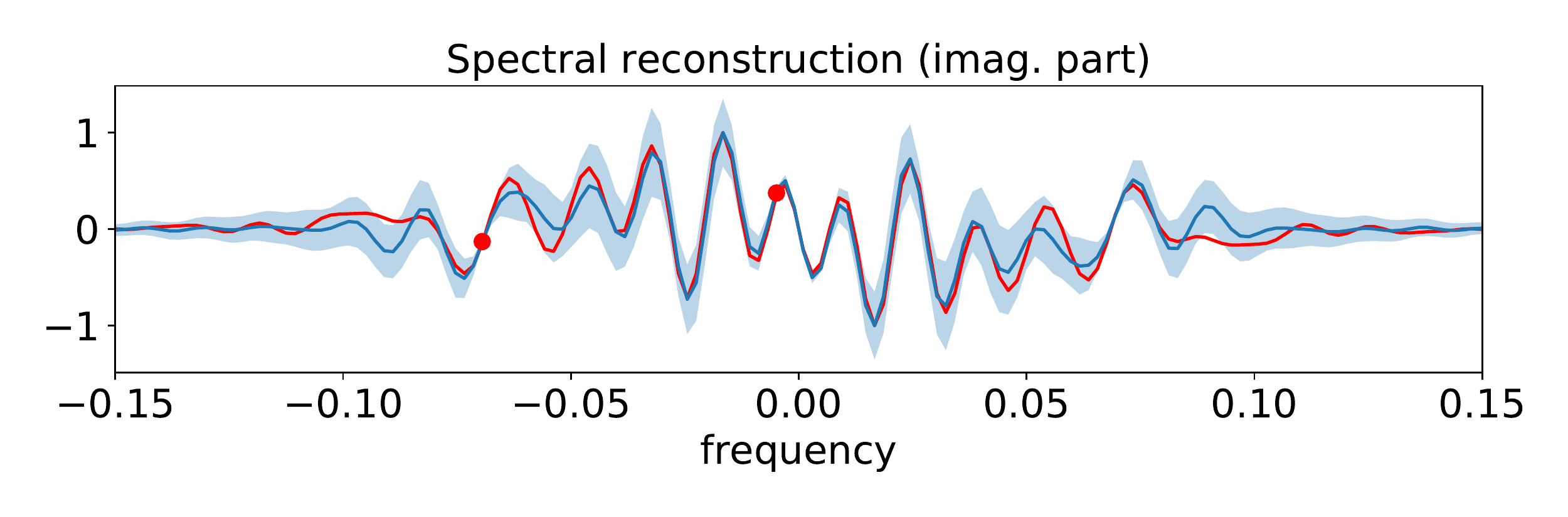}
    \end{subfigure}
    \caption{BRFP's spectral reconstruction of a synthetic time series using 2\% of noise-corrupted time and frequency observations estimation. The posterior mean is in solid blue and the 95\% confidence band is in light blue.} %\cyan{\bf[WERE THERE ONLY TWO OBSERVATIONS OF THE SPECTRA?]}}
    \label{fig:spectrum-estimation}
\end{figure}

Figs.~\ref{fig:time} and \ref{fig:spectrum-estimation} show the posterior  reconstructions of time and frequency respectively using BRFP. Notice that only with a limited number of observations, the posteriors are tight around the latent variables. A key feature of BRFP is that, despite the reduced amount of spectral observations, the posterior over the spectrum is concentrated around zero for high frequencies; this is because the (prior) SE covariance for the time series in eq.~\eqref{eq:SE_kernel} contains mainly low-frequency energy. 

Additionally, as discussed in the previous section, observe from Fig.~\ref{fig:spectrum-estimation} that the real (cf. imaginary) part of the spectrum is even (cf. odd). This contributes to an improved reconstruction, since knowing the value of the spectrum for a given frequency automatically provides information on the spectrum for its negative frequency counterpart.

\subsubsection{Identifying periodicities from temporal observations}

We considered the synthetic sum-of-sinewaves 
\begin{equation}
  \label{eq:synth_f}
  f(t) = 10\cos(2\pi 0.5t) - 5 \sin(2\pi t),
\end{equation}
for which we only observed 52 data points randomly selected across the range $[0, 10]$ and corrupted them with  white Gaussian noise with variance $\st^2 = 0.25$. Our aim was to use these partial/noisy/unevenly-sampled observations to estimate the fundamental frequencies $f_0 = 0.5$ and $f_1 = 1$ using the proposed BRFP and then compare these estimates against the ones of the classical Lomb-Scargle (LS) method. \revised{We remind the reader that LS, or LS periodogram, is a classical method for spectral estimation that is able to deal with irregularly sampled observations \cite{Lomb1976,Scargle1982}. The LS periodogram estimates the power spectral density by fitting a dictionary of sinusoidal bases to the data in a least squares sense. The dictionary, which is sampled at irregular times, is orthogonalised through a delay factor so that the basis is orthogonal and the fitting is unique.}

\begin{figure}[t]
    \centering
    \includegraphics[width =\linewidth]{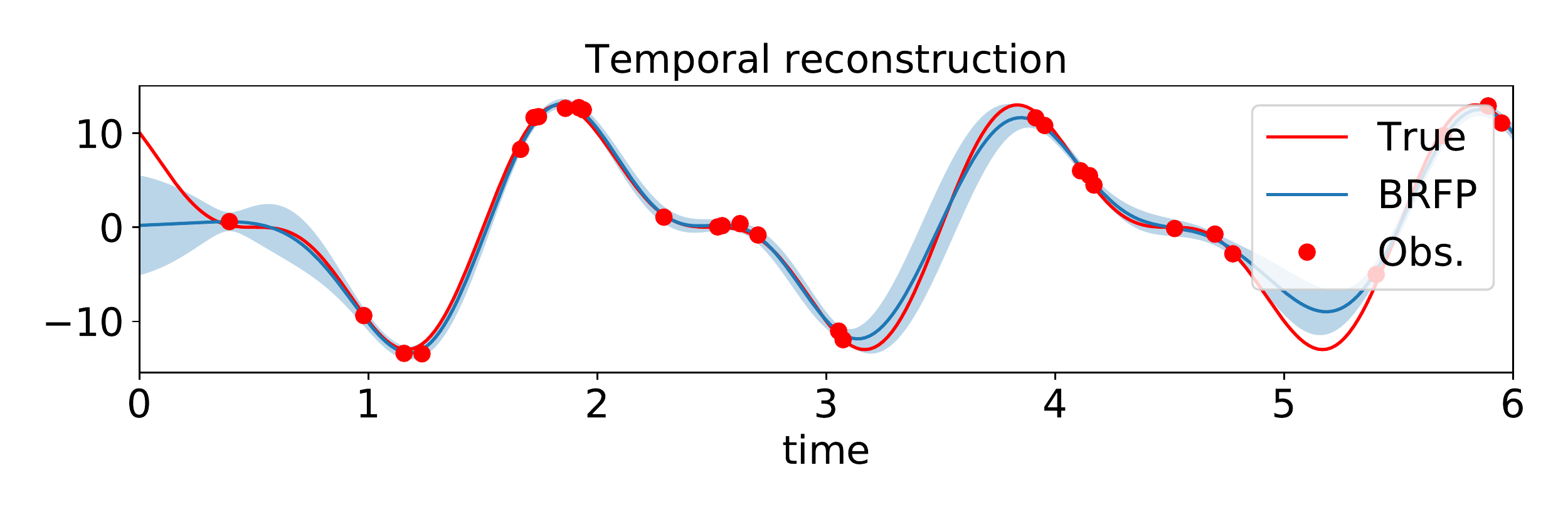}
    \caption{BRFP's temporal reconstruction of a sum-of-sinewaves, posterior mean in solid blue and 95\% confidence band in light blue.}
    \label{fig:time_periodogram}
\end{figure}

\begin{figure}[t]
    \centering
    \begin{subfigure}[t]{\linewidth}
        \includegraphics[width=\linewidth]{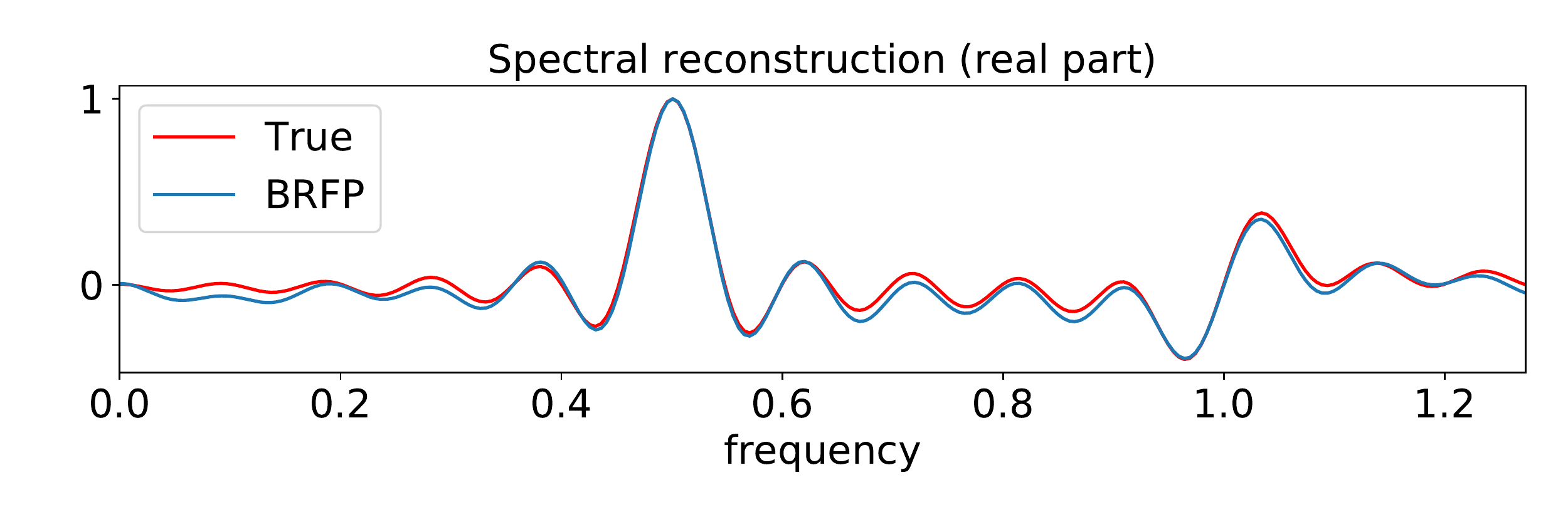}
    \end{subfigure}
    \begin{subfigure}[t]{\linewidth}
        \includegraphics[width=\linewidth]{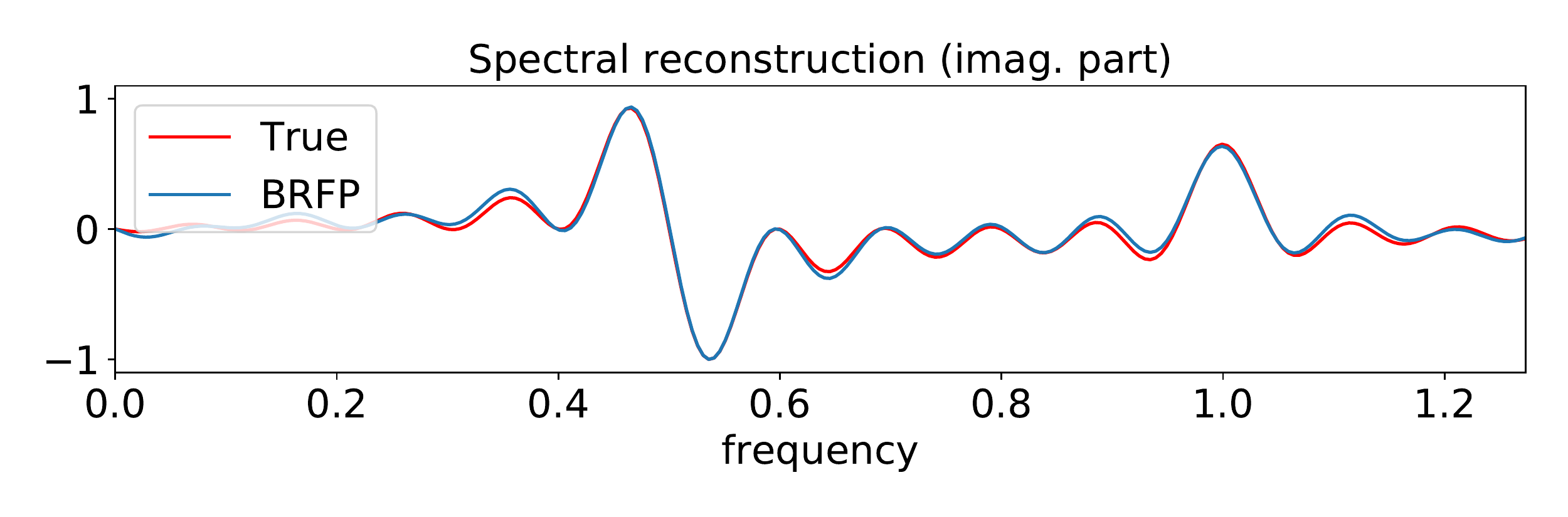}
    \end{subfigure}
    \begin{subfigure}[t]{\linewidth}
        \includegraphics[width=\linewidth]{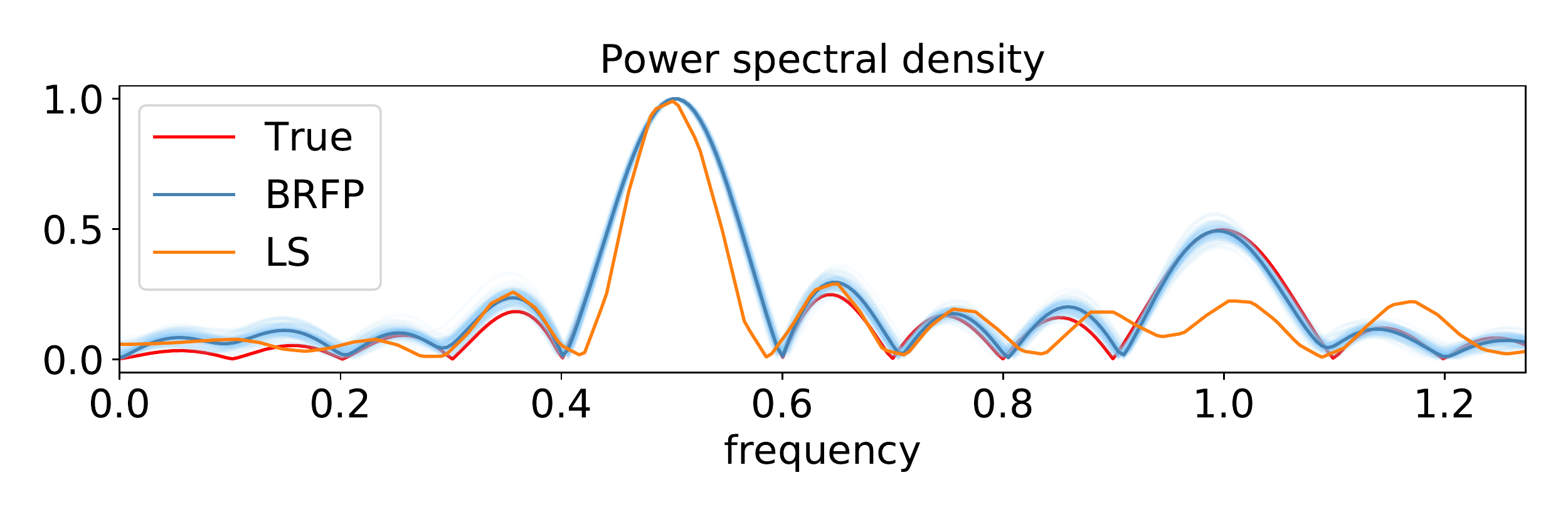}
    \end{subfigure}
    \caption{Spectral reconstruction of a sum-of-sinewaves only using temporal measurements. Lomb-Scargle directly computes the power spectrum and thus, the spectrum estimates are not available for comparison against the estimates of BRFP.}\label{fig:spectrum_periodogram}
\end{figure}

Fig.~\ref{fig:time_periodogram} shows the true function in eq.~\eqref{eq:synth_f} alongside the observations and the temporal reconstruction. BRFP was trained with an SE kernel and zero mean via maximum likelihood. Fig.~\ref{fig:spectrum_periodogram} shows the spectral reconstruction alongside the true spectrum computed via DFT over an evenly-sampled grid of $256$ points in the interval $[0, \frac{4}{\pi}]$. The figure shows the real and imaginary parts of the spectrum and the power spectrum at the bottom. We clarify that the power spectrum for BRFP was computed by adding square samples from the real and imaginary parts of its spectral reconstruction. Notice that the error bars of the proposed reconstruction in both the time and frequency domains are tight, despite the reduced quantity of temporal observations. The proposed method successfully identifies both energy peaks in frequency and resembles the complete power spectrum very closely. The LS benchmark, on the contrary, fails to follow the shape of the power spectrum and provides poor estimates of the power spectrum around the frequency of the second sinusoid. %, specially those of the two main lobes.

%!TEX root = ./BRFP.tex

\section{Experiments}
\label{sec:simulations}
This section validates BRFP on three experimental settings against both classic and modern benchmarks. The first experiment (E1) focuses on estimating the spectrum of an incomplete and noise-corrupted heart-rate time series in order to detect a specific harmonic component. The second experiment (E2) evaluates the robustness of BRFP with respect to increasing levels of observation noise and missing data; this setting considered an audio recording and benchmarks BRFP against classic and modern spectral estimation methods. The third experiment (E3) tests BRFP for two-dimensional data via the reconstruction of an astronomical image from corrupted/partial 2D spectral (interferometry) measurements. All benchmark methods against which BRFP are compared are briefly presented on Appendix \ref{sec:benchmarks}, while the 2D extension of BRFP, required for experiment E3, is explained on Appendix \ref{sec:2dmodel}.

The next section presents the metrics for quantitative assessment to be used across our experiments. Then, Sections  \ref{sec:E1}, \ref{sec:E2} and \ref{sec:simulations_alma} show experiments E1, E2 and E3, respectively.

\subsection{Performance metrics} 

Both temporal and spectral estimates will be assessed by means of the normalised mean square error (NMSE), which is the square ratio between the $\ell_2$ norm of the estimate error and the $\ell_2$ norm of the ground truth, that is, 
\begin{equation}
    \text{NMSE} = \frac{\sum_{i=1}^N (x_i-\hat{x}_i)^2}{\sum_{i=1}^N x_i^2} ,
\end{equation}
where $x_i$ and $\hat{x}_i$ are the $i$-th value of the true and estimated quantity respectively. The choice of the NMSE stems from the fact that the generative model is Gaussian and therefore the sum of squares is the empirical variance, the natural measure of dispersion.

When it comes to  \emph{power spectrum}, that is, the square magnitude of the spectrum $p_k = [\Xr]_k^2 + [\Xi]_k^2$, quantitative comparison is far more challenging and a matter of investigation in itself \cite{flamary2016optimal,Wasserstein-Fourier}. This is because the standard $\ell_2$ norm quadratically penalises large deviations thus being insensitive to small inaccuracies which are fundamental when estimating energy at, e.g., harmonics. In this sense, we consider two metrics to properly compare an estimate PSD $\hat{p}$ with a ground truth PSD $p$. The first one will be the $\ell_{0.1}$ norm given by \begin{equation}
   \ell_{0.1}(p-\hat{p}) = \frac{1}{N}\left(\sum_{k=0}^{N-1} |p_k-\hat{p_k}|^{0.1}\right)^{10},
\end{equation}
which has been chosen, instead of say the $\ell_2$ norm, to give emphasis to small deviations. Intuitively, this norm resembles the idea of whether an estimate \emph{deviates from the ground truth or not}, rather than \emph{how much it deviates}. Notice that by taking an even smaller exponent, we would have the $\ell_0$ norm, which is a count of incorrect estimates, widely used in compressed sensing.

The second metric for comparison of PSDs will be the Kullback-Leibler (KL) divergence defined by
\begin{equation}
    \text{D}_{\text{KL}}\left(p\|\hat{p}\right) = \sum_{k=0}^{N-1} p_k\log\left(\frac{p_k}{\hat{p}_k}\right),
\end{equation}
which requires that both $p$ and $\hat{p}$ integrate to one, meaning that the KL divergence measures the relative spectral energy spread rather than its magnitude. The KL divergence is an information-theoretic measure that quantifies how well $\hat{p}_k$ represents the information in $p_k$, which is exactly what we require. In particular, notice that if $p_k>0$ and $\hat{p}_k=0$, then the KL becomes infinite, meaning that we will \emph{infinitely penalise} PSD estimates $\hat{p}$ that fail to convey energy at frequencies that do contain energy under the reference PSD $p$.

\subsection{E1: Detecting critical harmonic activity (heart-rate signal)}
\label{sec:E1}
\begin{figure}[t]  
    \centering
    \includegraphics[width =\linewidth]{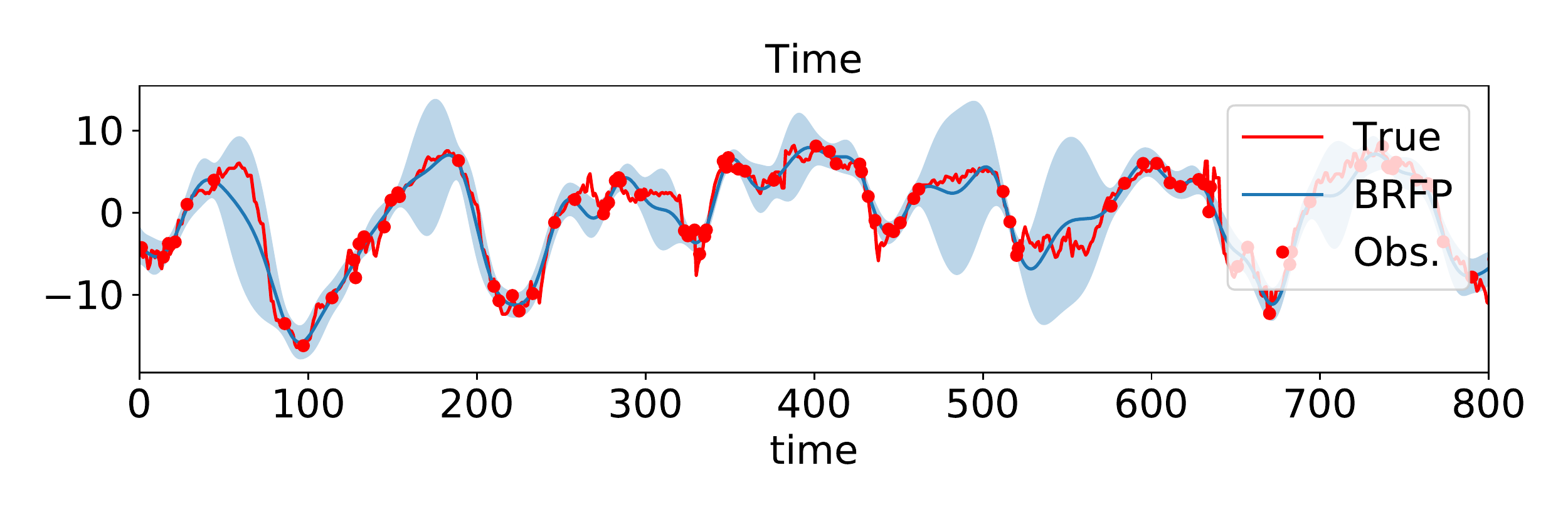}
    \caption{BRFP's temporal reconstruction of a real-world heart-rate time series. The posterior mean is in solid blue and the 95\% confidence band is in light blue.}
    \label{fig:time_periodogram2}
\end{figure}

\begin{figure}[t]
    \centering
    \begin{subfigure}[t]{1\linewidth}
        \includegraphics[width=\linewidth]{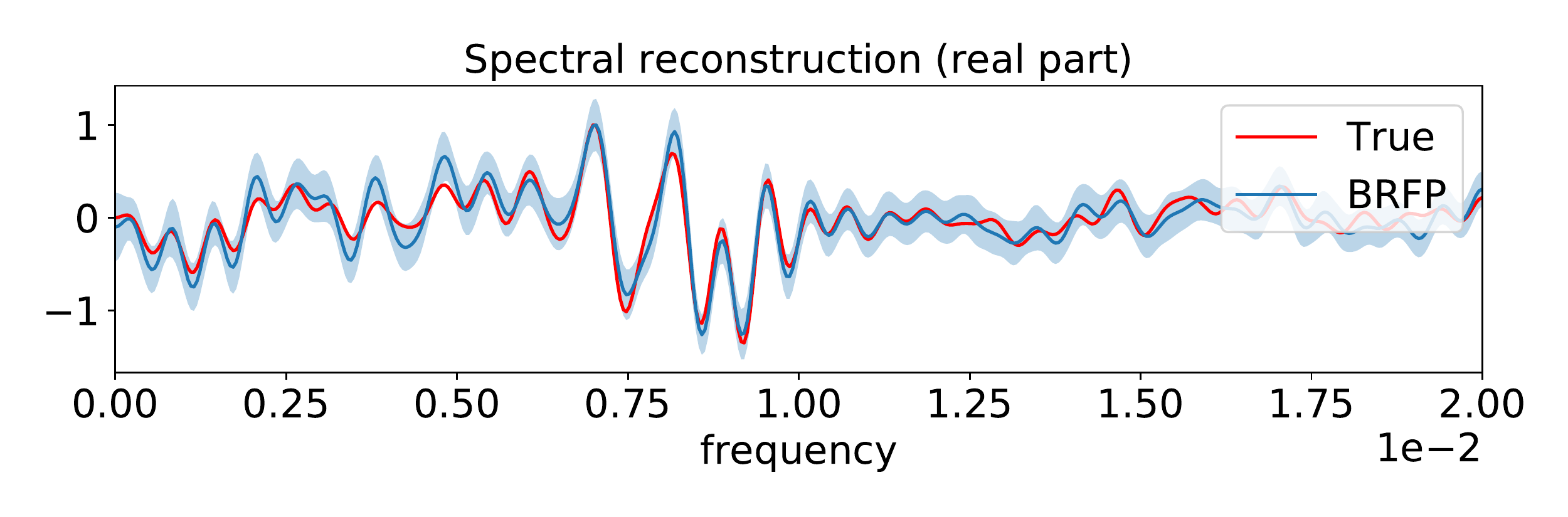}
    \end{subfigure}
     \begin{subfigure}[t]{1\linewidth}
         \includegraphics[width=\linewidth]{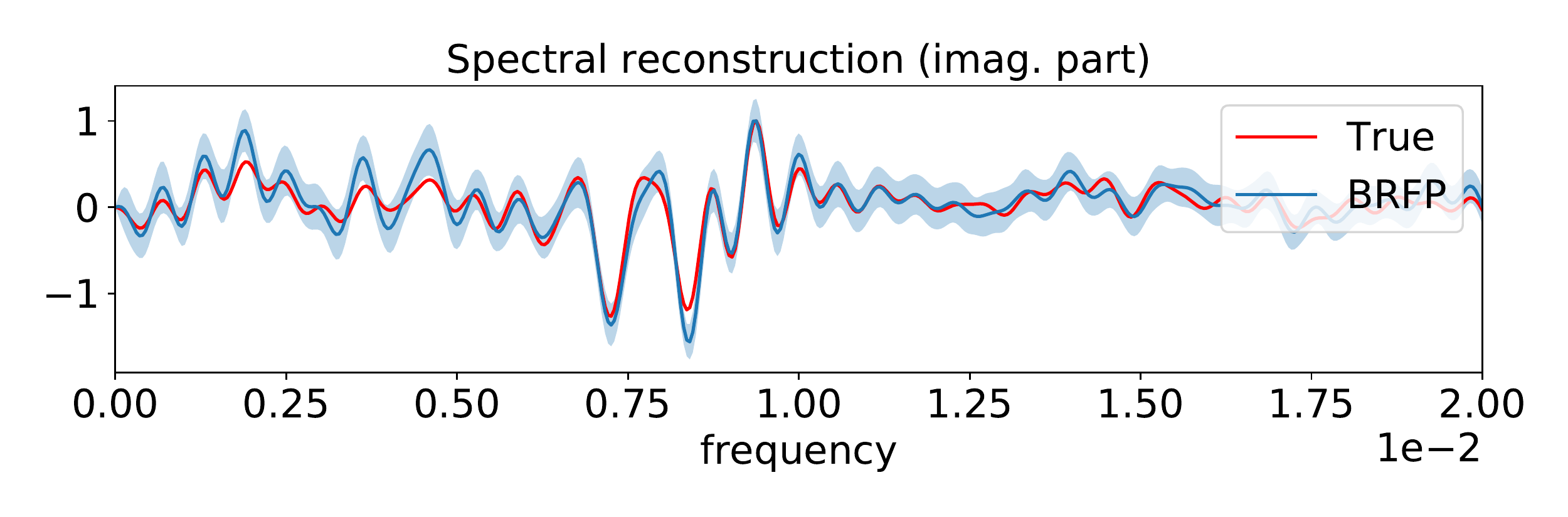}
     \end{subfigure}
    \begin{subfigure}[t]{1\linewidth}
        \includegraphics[width=\linewidth]{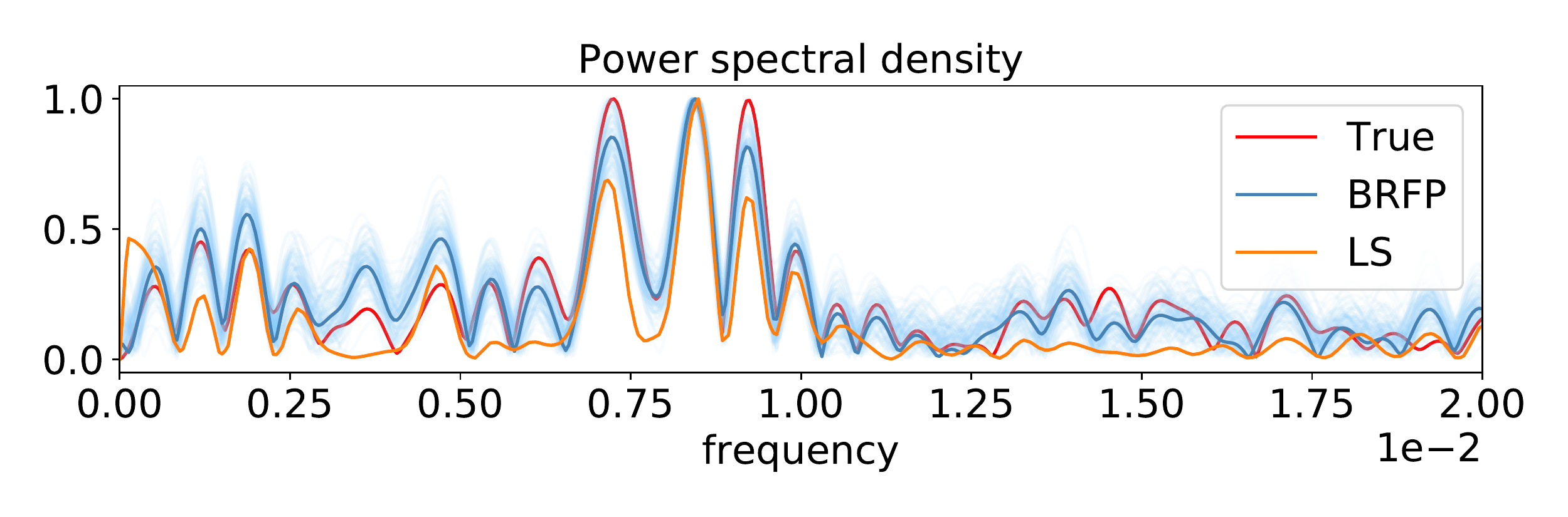}
    \end{subfigure}
    \caption{Spectral reconstruction of a heart-rate series using temporal measurements: BRFP (SE kernel) and Lomb-Scargle. The BRFP's posterior mean is in solid blue and the 95\% confidence band is in light blue. }\label{fig:spectrum_periodogram2}
\end{figure}

We considered a heart-rate (HR) recording from the MIT-BIH database.\footnote{Source: \url{http://ecg.mit.edu/time-series/}.} This signal corresponds to a patient with congestive heart failure, a chronic condition that affects the heart muscles \cite{Goldberger1991}. This condition, which is characterised by a HR with low-frequency activity, can be detected via electrocardiogram. In this sense, our aim is to detect low-frequency spectral energy from the time series using only partial/noisy observations that simulate a real-world recording. Therefore, we considered an uneven sampling of only  10\% of the available data corrupted by white Gaussian noise with $\st^2 = 0.25$. Figs.~\ref{fig:time_periodogram2} and \ref{fig:spectrum_periodogram2} show the temporal and spectral representation respectively, comparing the true signals, the observations and the BRFP reconstruction against that of LS---see Appendix \ref{sec:benchmarks}.

Observe from Fig.~\ref{fig:time_periodogram2} that the reconstruction of the temporal representation exhibits some inaccuracies in the form of wide error bars in the regions of missing data---see e.g., time indices around 500. This is a consequence of the Gaussian assumption, since this signal comprises a pseudo-periodic waveform with wide plateaus and, thus, it is hardly Gaussian. We considered this challenging setting to show that BRFP is robust to model misspecification: the uncertainty associated to the wide error bars in time is integrated out when it comes to find the posterior spectral representation. This is evident from Fig.~\ref{fig:spectrum_periodogram2}, where we can see that the three peaks just below frequency 0.01---which characterise the congestive heart failure---are sharp and well defined by BRFP. 


\subsection{E2: Robust spectral estimation (audio)}
\label{sec:E2}
\begin{figure}[t]
    \centering
        \includegraphics[width=\linewidth]{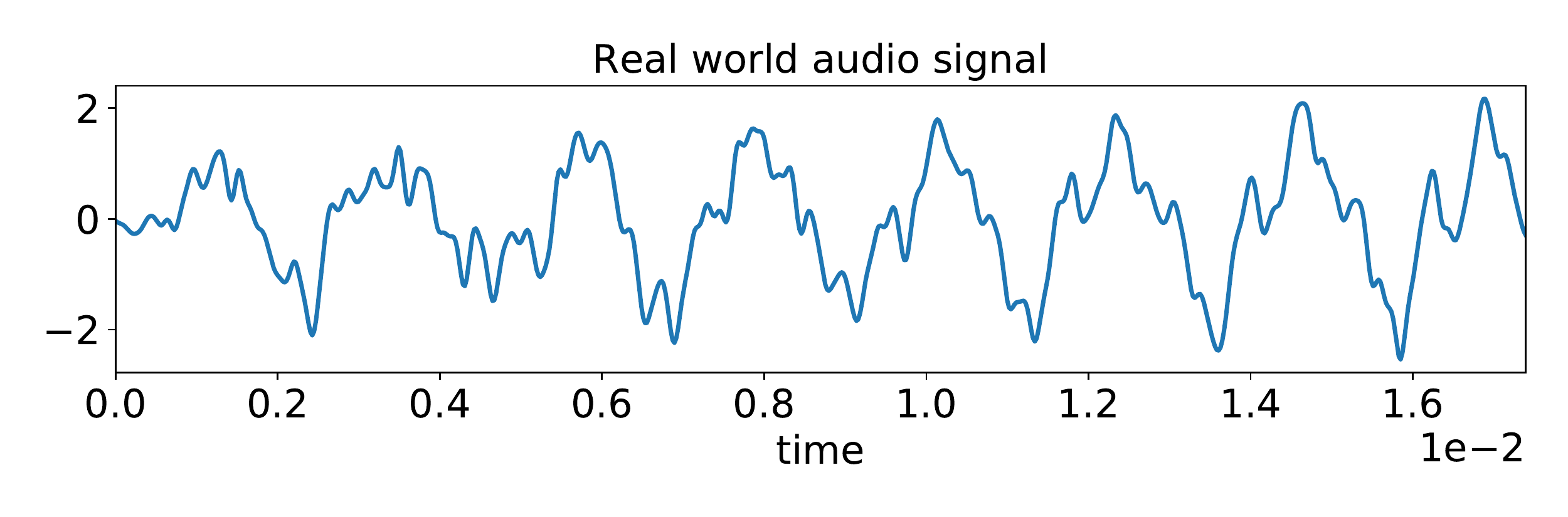}
        \caption{Note \emph{A} (or \emph{La}), with a frequency of 440Hz, played by a classical piano. The sampling frequency was 44.1kHz}
        \label{fig:piano}
\end{figure}

% sigma**2*np.exp(-alpha*(np.sin(beta*np.abs(x1-x2)) ** 2))
The second experiment evaluated BRFP's robustness to different levels of observation noise and missing data. To this end, we considered an audio signal corresponding to the note \emph{A} (or \emph{La}), with fundamental frequency at 440Hz, played by a classical piano.\footnote{Source: \url{https://lmms.io/lsp/?action=show&file=4346}} The recording---shown in Fig.~\ref{fig:piano}---was taken at a sampling rate of 44.1kHz and comprised $N=768$ evenly-sampled observations. Since this audio signal features a strong periodic component, we considered the periodic covariance kernel for BRFP, proposed by \cite{mackay_gp}, given by
\begin{equation}
K(t,t') = \sigma^2\exp(-\alpha\sin^2(\beta|t-t'|)),
\end{equation}
where the hyperparameters denote the marginal variance ($\sigma^2$), the rate ($\alpha$) and the frequency ($\beta$).

\revised{This experiment comprises two parts, namely the Noise Test (where BRFP is tested against classical methods) and the Missing-Data Test (where modern benchmarks are considered) to be presented as follows. All benchmarks are   described on Appendix \ref{sec:benchmarks}}

\subsubsection{Noise Test} \label{sec:noise_test} We first analysed the entire recording (no missing data) and added white Gaussian noise with increasing magnitude to compare the proposed BRFP to LS, Welch  (using non-overlapping Hanning windows of size $384$) and YW (with AR order of $400$). We chose $12$ values for the noise standard deviation evenly-spaced in the range $\sigma \in [0.1,1.2]$. For all considered models and levels of noise, we computed the KL divergence and the $\ell_{0.1}$ distance against the ground-truth power spectrum (given by the Periodogram using the complete noiseless signal) over 10 realisations, thus being able to compute error bars. 

\begin{figure}[t]
\centering
        \includegraphics[width=0.7\linewidth]{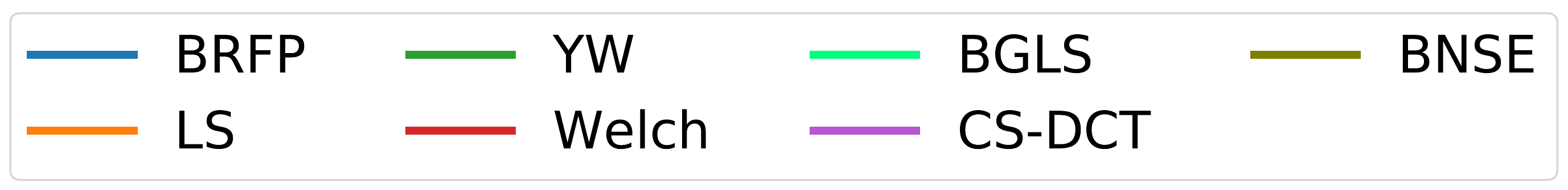}
        
        \includegraphics[width=0.49\linewidth]{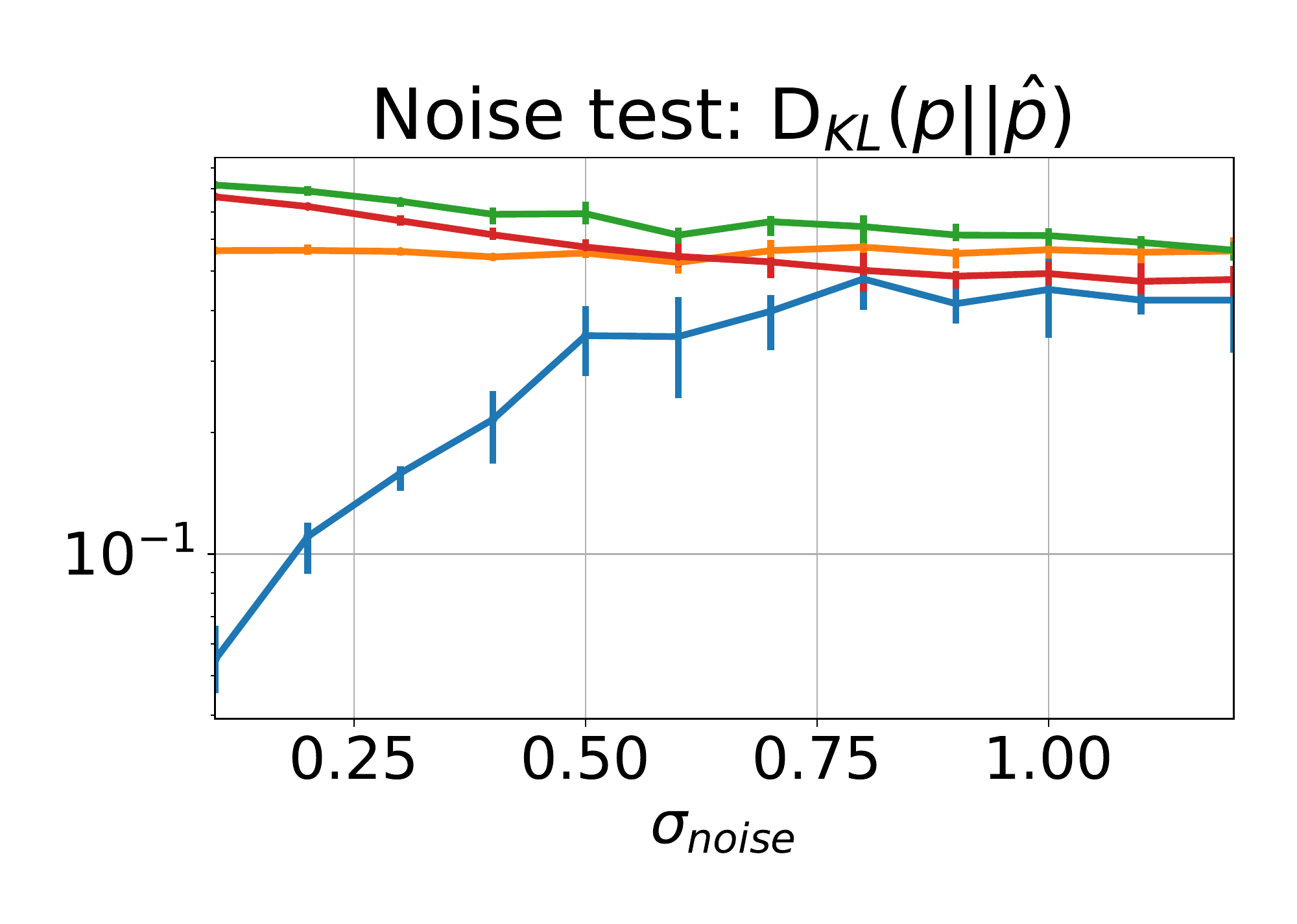}
        \includegraphics[width=0.49\linewidth]{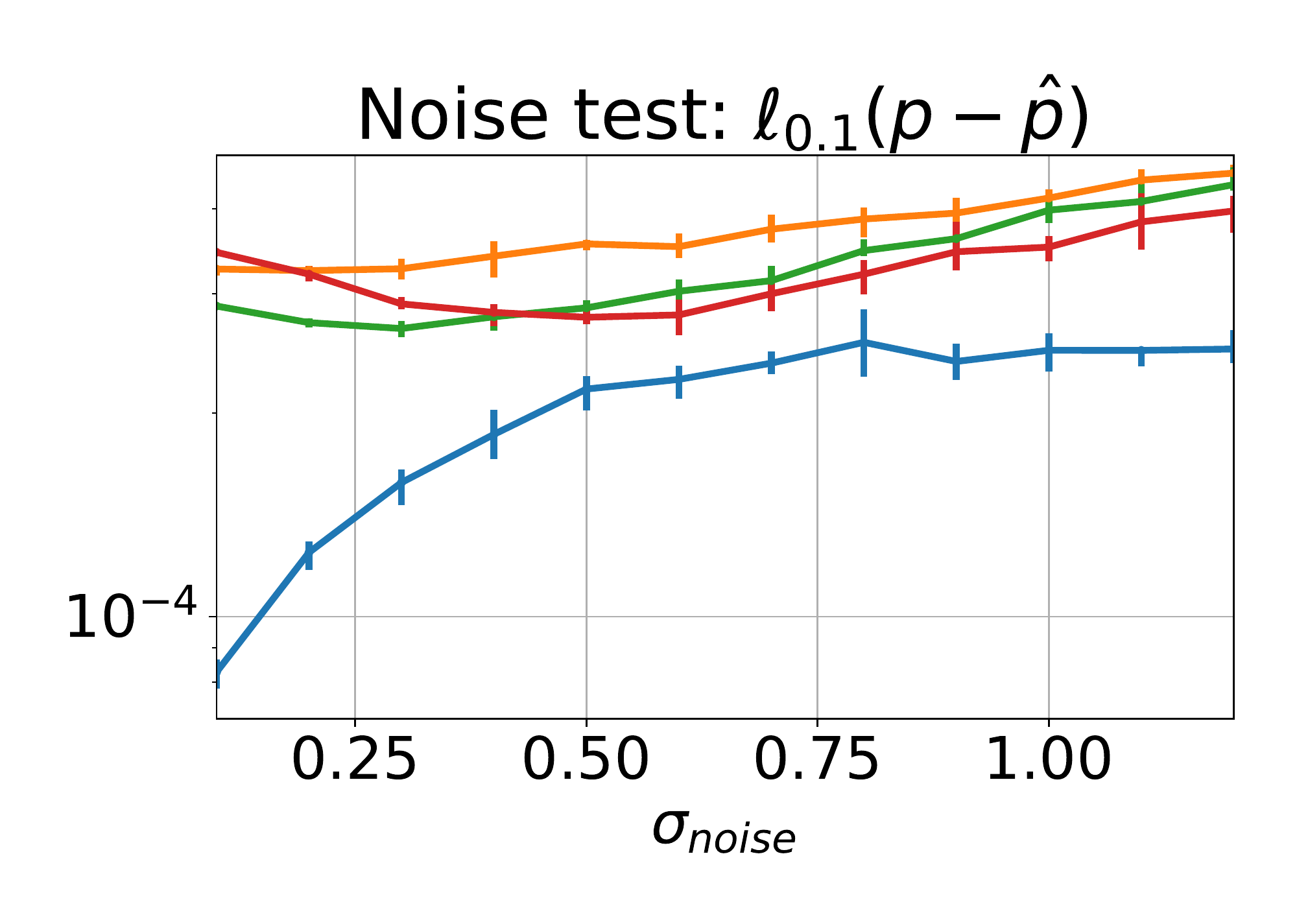}
        \includegraphics[width=0.49\linewidth]{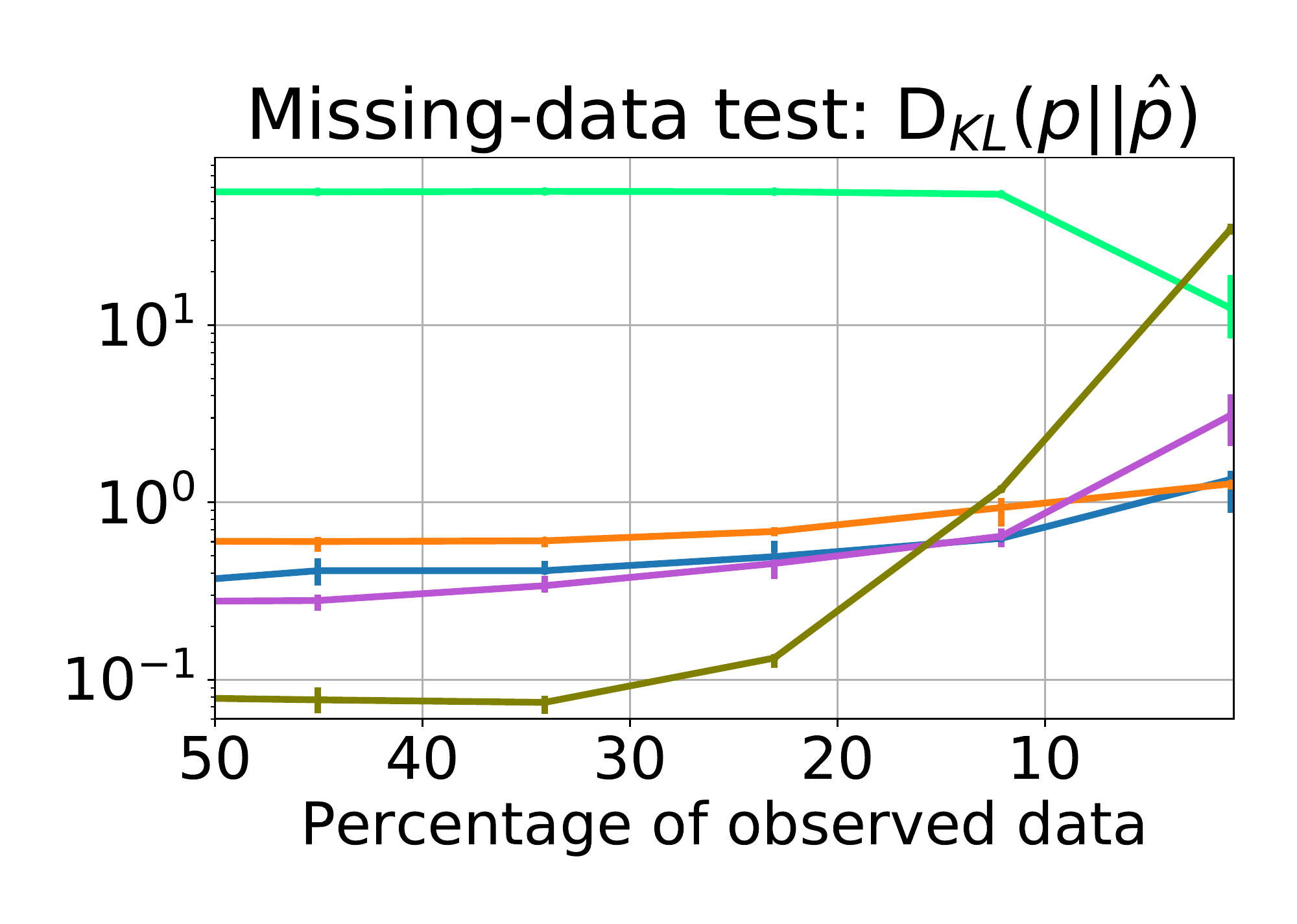}
        \includegraphics[width=0.49\linewidth]{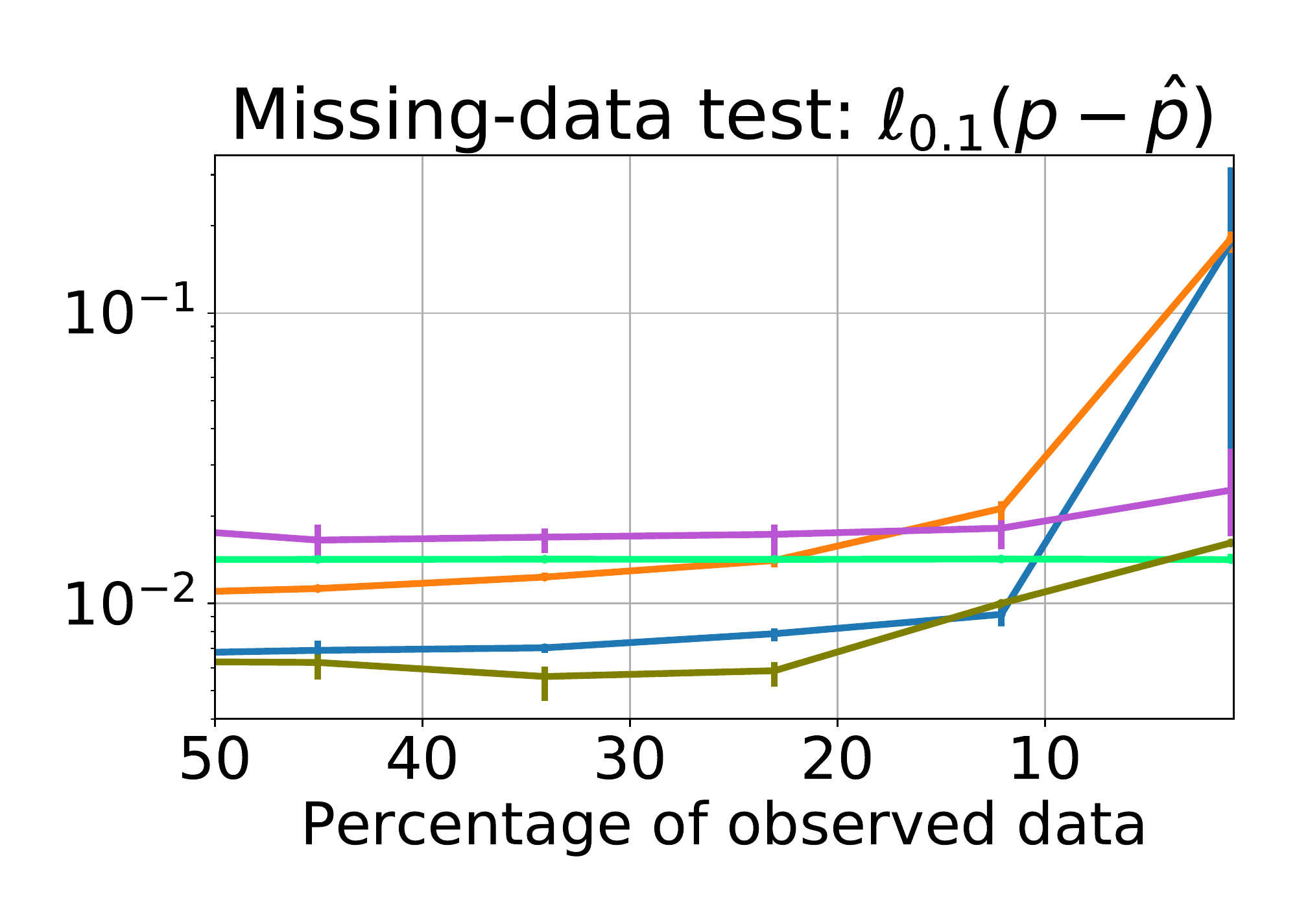}
    \caption{Robustness  of  spectral  estimates. The top and bottom rows represent the noise and missing-data tests in Secs.~\ref{sec:noise_test} and \ref{sec:missing_test} respectively. The left and right columns show the KL and the $\ell_{0.1}$ metrics respectively. All error  bars  correspond to the 20 and 80 percentile over 10 independent realisations.}\label{fig:audio_tests}
\end{figure}

Fig.~\ref{fig:audio_tests} (top) shows the performance of the considered methods in this test, where BRFP outperformed the classic methods for all levels of noise and both metrics. Notice that, under KL (top-left), the benchmarks seem to slightly improve with noise, which is due to the normalisation step that neglects the error in estimating the low-magnitude harmonics. On the contrary, the $\ell_{0.1}$ norm (top-right) focuses on the amount of small errors, mainly in the harmonics, where the gap between BRFP and the benchmarks is clear.

\subsubsection{Missing-Data Test}
\label{sec:missing_test}
We then randomly (uniformly) removed a portion of the data in the range $[50\%,99\%]$ and implemented BRFP against LS, and the modern methods for spectral estimation BGLS, CS-DCT and BNSE (see Appendix \ref{sec:benchmarks} for acronyms) for non-uniformly-sampled data. This experiment began implementing the above methods with $384$ observations (half of the recording data) and concluded with only $8$ observations; i.e., with 99\% of missing data. For a realistic scenario, this experiment also considered observation noise with fixed standard deviation $\sigma = 0.5$, and was also repeated 10 times, since the data removal process is random, thus obtaining error bars. Fig.~\ref{fig:audio_tests} (bottom) shows the results of this test, where BRFP and BNSE are the competitive methods.  
\begin{figure}[t]
    \centering
        \includegraphics[width=\linewidth]{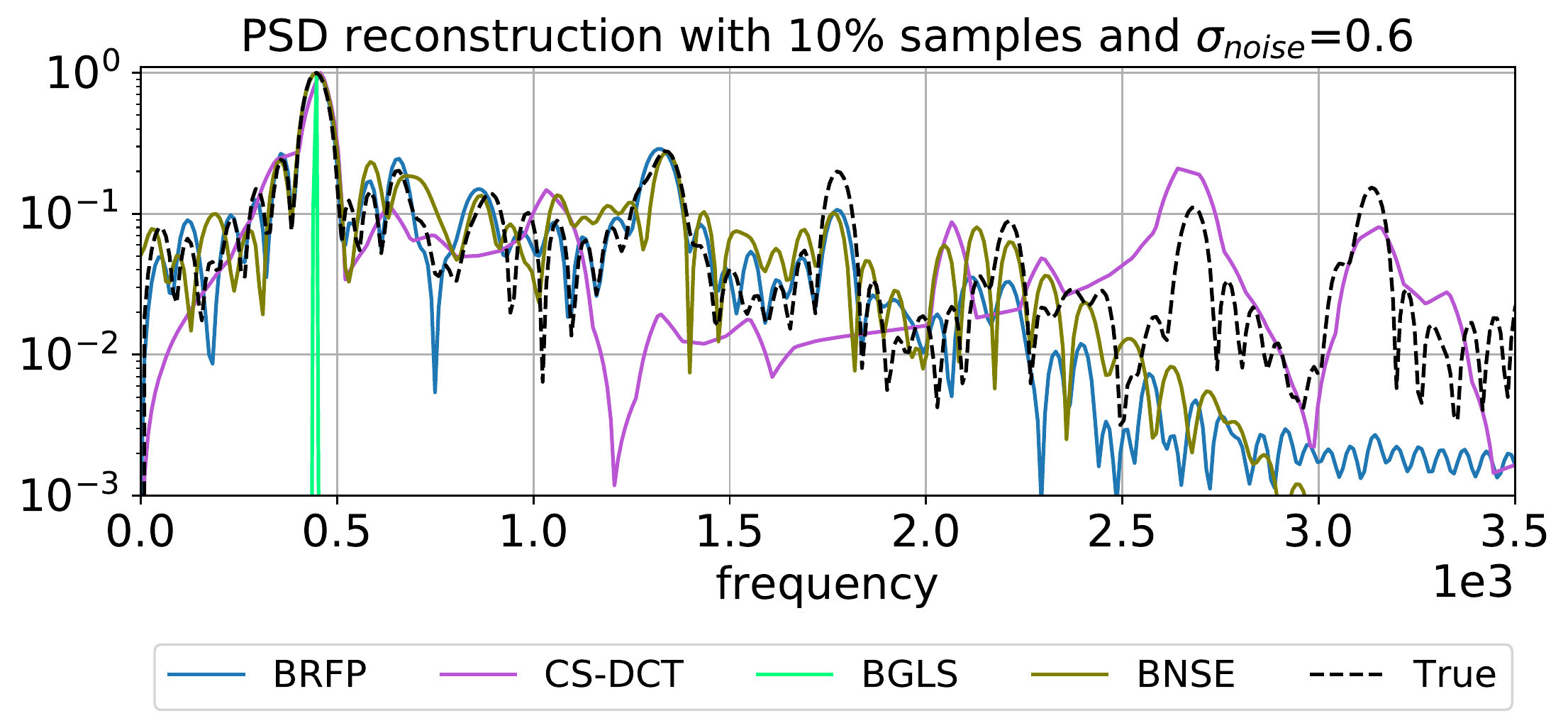}
        \caption{PSD estimates of the piano audio using BRFP and other benchmarks.}
        \label{fig:piano_modern}
\end{figure}

\revised{
For both tests and performance metrics, we emphasise that tight error bars (for the 20-80 percentiles) were achieved only with 10 independent realisations: this allows us to identify a clear statistical significance of the difference between the performance of all the considered methods. Additionally, and for illustrative purposes, we include Fig.~\ref{fig:piano_modern} which shows the actual estimates of the PSD using BRFP and the benchmarks in the missing data test (10\% of samples, 0.6 noise std). Notice how BRFP faithfully follows the true PSD up to frequency $\approx 1.7$, together with BNSE, while the other methods fail to provide sound estimates.  Critically, notice how the LS variant can only find the main periodicity but fails when it comes to represent the entire PSD. The similarity in performance for both the proposed BRFP and BNSE stems from their common Gaussian structural assumption, yet the advantage of the BRFP is a practical one discussed on Section \ref{sec:discussion}.
}

\subsection{E3: From spectral to spatial domain (interferometry)}
\label{sec:simulations_alma}

Radio interferometry \cite{Thompson2017} is a technique in which signals from widely-spaced antennas are combined to produce high-resolution images of cold and/or obscured structures of our Universe, such as planet-forming regions around young stars or the event horizon of a black hole.  

\begin{figure}[t]
	\centering
		\makebox[\linewidth]{\includegraphics[trim={3cm 2.5cm 2.3cm 3cm},clip, width=\linewidth]{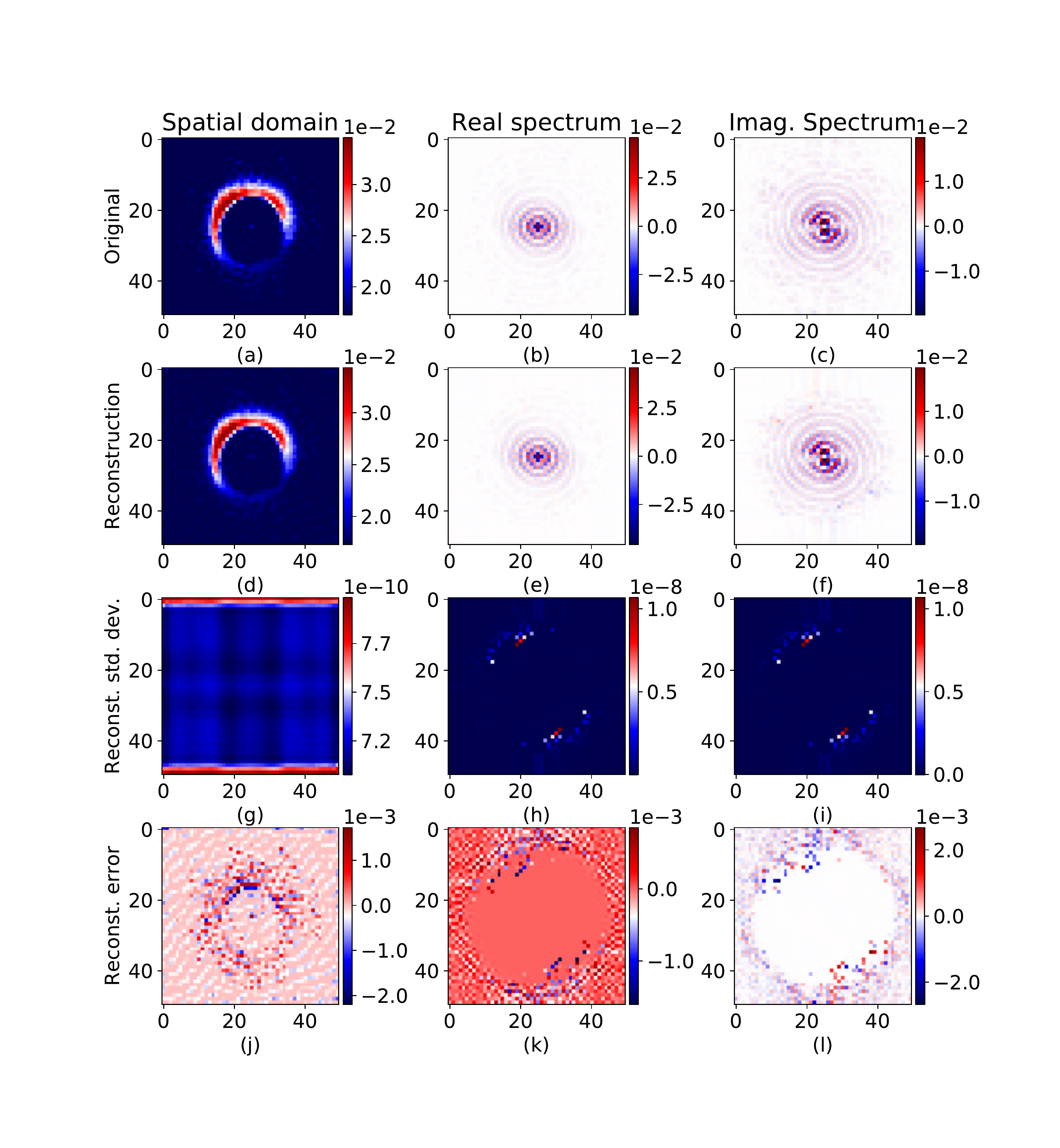}}
		\caption{Reconstruction of HD142527 using BRFP. From left to right: spatial domain, real spectrum, imaginary spectrum. From top to bottom: Original image, BRFP reconstruction, reconstruction st.dev and reconstruction error.}
		\label{fig:BSR-2D-54data}
\end{figure}

The observations of a radio interferometer correspond to a sparse and irregularly-sampled set of the 2D Fourier spectrum of the radio source. As direct application of the inverse Fourier transform is not possible, the radio image is typically reconstructed using the CLEAN heuristic \cite{CLEAN}. Fig.~\ref{fig:BSR-2D-54data} (top, left) shows a 50-by-50 pixel image of young star HD142527 featuring a protoplanetary disk reconstructed with CLEAN, from spectra observed by \cite{Casassus2015accretion} through the ALMA observatory.\footnote{Dataset ALMA ``2011.0.00465.S'', see the appendix of \cite{Casassus2015accretion} for more details.} 

In what follows, we apply BRFP to reconstruct the radio-image of young star HD142527. This experiment aims to i) show that the proposed BRFP can reconstruct HD142527 from incomplete spectral measurements, that is, to provide an image close to the CLEAN reconstruction (our benchmark), while also being able to ii) place error bars for the reconstruction, thus providing an advantage to existing reconstruction methods. %In order that the proposed BRFP admits 2D data, we incorporated the transformation described in \ref{sec:2dmodel}.
The practical generalisation of the BRFP formulation to 2D data can be found in Appendix \ref{sec:2dmodel}.

\begin{figure}[t]
	\centering
	\begin{minipage}[b]{0.45\linewidth}
		\includegraphics[width=\linewidth]{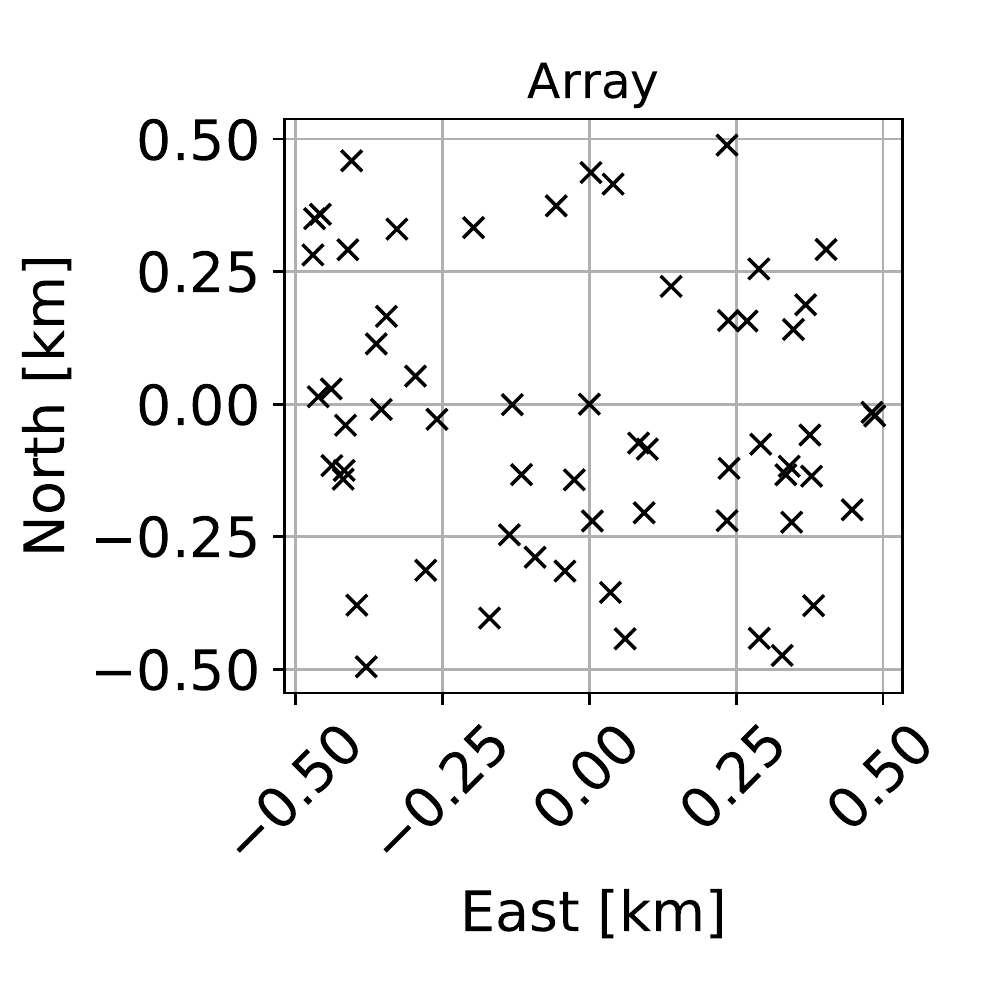}
    \end{minipage}
	\begin{minipage}[b]{0.45\linewidth}
        \includegraphics[width=\linewidth]{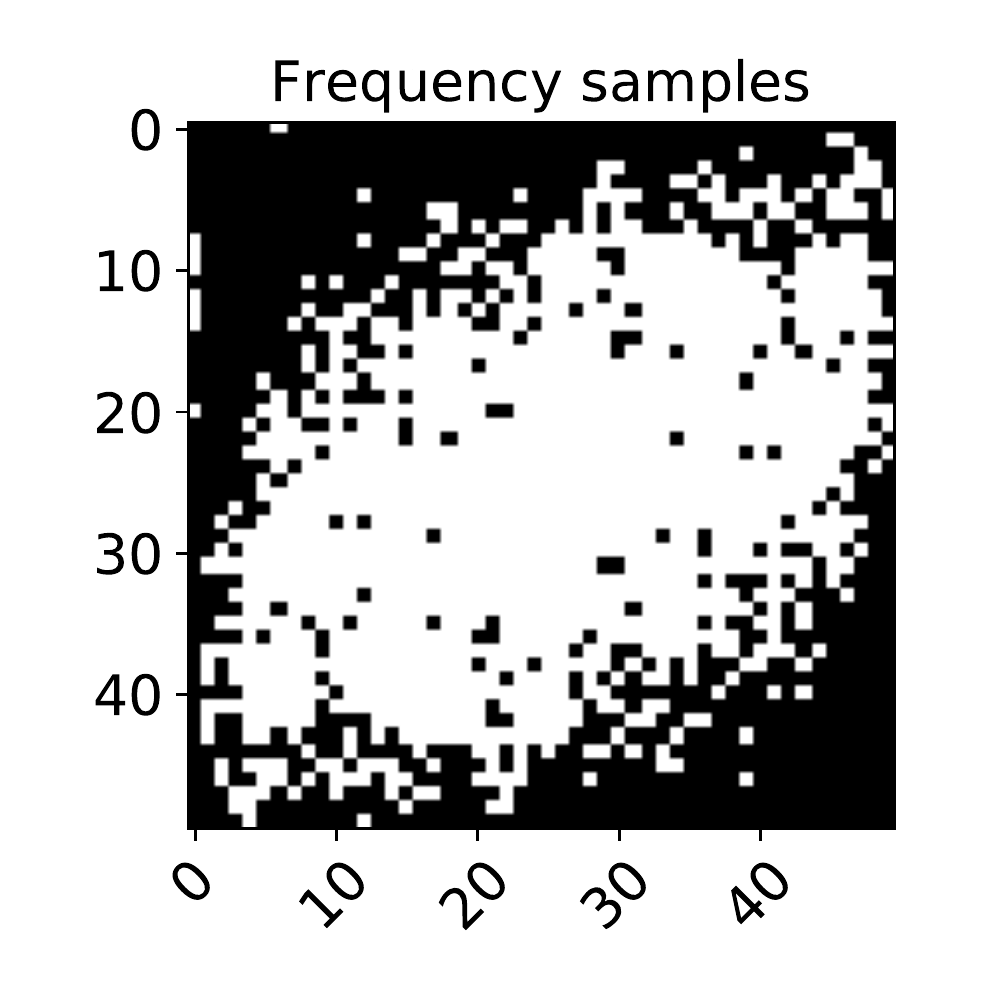}
    \end{minipage}

    \caption{Layout of antennas and sampling frequencies considered for the reconstruction of HD142527. Left: Array of antennas, where each cross represents an antenna in the interferometry array. Right: spectral observation pattern, where white pixels denote observed frequencies and black pixels denote unobserved ones. This array features 60 antennas and it covers 54\% of the (discrete-space) frequencies.}
    \label{fig:antennas_layout}
\end{figure}

The frequencies observed in an interferometry application are determined by the spatial layout of the array of antennas. For our experiment, Fig.~\ref{fig:antennas_layout} shows the considered layout of antennas (left) and the corresponding spectral sampling pattern (right) that such a layout induces on the frequency space.

\begin{table}[t]
\centering
\begin{tabular}{l|c}
\hline
                        & \textbf{NMSE (for BRFP)}          \\ \hline
\textbf{Spatial domain} & $1.17  \cdot 10 ^{-3}$ \\
\textbf{Real spectrum}  & $5.92  \cdot 10 ^{-5}$ \\
\textbf{Imag. spectrum} & $1.69  \cdot 10 ^{-2}$ \\ \hline
\end{tabular}
\caption{NMSE for interferometry reconstruction in each domain using BRFP.}
\label{tab:interferoetry-errors}
\end{table}

Table \ref{tab:interferoetry-errors} shows the NMSE of the BRFP reconstruction for both representations. Fig.~\ref{fig:BSR-2D-54data} shows the reconstruction images using BRFP, where both for the spatial and spectral representations the ground truth (CLEAN) is presented alongside BRFP's the reconstruction error and standard deviation. 

First of all, notice that the spatial reconstruction provided by BRFP, Fig.~\ref{fig:BSR-2D-54data} (d), is visually indistinguishable from the ground truth, Fig.~\ref{fig:BSR-2D-54data} (a). In fact, notice from Fig.~\ref{fig:BSR-2D-54data} (j) that the reconstruction error is one order of magnitude smaller that the images (with overall NMSE =$1.17 \cdot 10^{-3}$). Additionally, observe that the reconstruction error, Fig.~\ref{fig:BSR-2D-54data} (j), takes values symmetrically wrt zero, suggesting that the spatial reconstruction is unbiased. Another key result is that, since the observations were acquired in the spectral domain, the standard deviation of the spectral reconstruction collapses to zero in the observed regions: compare Fig.~\ref{fig:antennas_layout} (right) to Figs.~\ref{fig:BSR-2D-54data} (k) and (l) to notice that the discrepancy between the spectral ground truth and BRFP reconstructions only occurs on the borders of the sampling pattern.

%!TEX root = ./BRFP.tex

\section{Discussion of results}
\label{sec:discussion}

The experimental ability of BRFP has been shown through examples on synthetic datasets and also real-world data from healthcare, audio and astronomy. \revised{The general take-home message of the experimental section is that BRFP performed successfully   for all settings and metrics, particularly for the missing-data test in Sec.~\ref{sec:missing_test}, where it outperformed most benchmarks and behaved similar to BNSE\footnote{BNSE stands for Bayesian Nonparametric Spectral Estimation; see Appendix \ref{sec:benchmarks} for its brief description.}. We argue that the reason for improved performance stems from the direct computation of the Fourier transform, a unique feature only present on BRFP and BNSE. Intuitively, both these methods analyse  incomplete/noisy data to perform interpolation and cleaning in a probabilistic manner, only then BRFP and BNSE compute the Fourier transform (or its inverse) of the resulting estimate, thus yielding a full probabilistic representation in time and frequency. In the graphical models jargon, we could say that both these methods \emph{integrate over} the posterior latent signal to compute the spectrum from the temporal observations and vice versa, rather than using the observations to explicitly compute  a point estimate of the time series, to then calculate the spectrum of such series. 

To emphasise the differences between the proposed BRFP and the recently introduced BNSE \cite{TobarNIPS2018}, we clarify that BNSE is a fully non-parametric method, which is only defined for windowed data due to the necessity of all trajectories being Lebesgue integrable (see Sec.~3.2 in \cite{TobarNIPS2018}). Therefore, the hypothesis of BNSE requires that the data come from a latent, infinite-dimensional but integrable, object. The proposed BRFP, on the contrary, is defined for discrete data and can be straightforwardly applied to it without violating any structural assumption. 

Additionally, another key feature of BRFP  against all methods considered (including BNSE) is that it can deal with inference in both ways: from time measurements to spectrum estimates, and from spectral observations to temporal estimates. The existing methods, conversely, are purpose specific and designed for only one of these ``directions''. 
}

The results of the presented experiments suggest that the proposed methodology provides accurate spectral analysis by a modern Bayesian treatment of the observed data. To the best of our knowledge, BRFP is the first method that
\begin{enumerate}
 	\item takes advantage of the fact that Gaussian signals have Gaussian Fourier spectra, and thus they can be estimated jointly;
 	\item models naturally the spectrum phase by considering both real and imaginary parts, unlike most of existing methods which only focus on the power spectrum magnitude; and
 	\item incorporates a flexible conjugate Gaussian prior and updates it into a posterior, thus inherently and efficiently modelling uncertainty in closed form. 
 \end{enumerate}

\section{Concluding remarks}
\label{sec:conclusion}

We have proposed \emph{Bayesian reconstruction of Fourier pairs} (BRFP), a novel methodology for joint reconstruction of temporal and spectral representations using partial observations. The proposed methodology focuses on discrete-time stationary signals and assumes a flexible Gaussian generative model for time series, together with a likelihood that allows for noise-corrupted observations and missing data. We have also analysed BRFP in theoretical terms against the classical spectral estimation perspective and have showed that BRFP can be used for general tasks in Fourier analysis such as spectral estimation, spatial reconstruction and periodicity detection. Both experimentally and conceptually, we have illustrated that BRFP collapses to standard methods (such as simply applying the DFT) when the usual assumptions of noise-free observations hold; however, BRFP also  equips the spectral analysis with a natural treatment of missing data and observation noise. 

\revised{
Besides its sound presentation in connection with both the spectral estimation and inference perspectives, BRFP has exhibited an appealing performance for modern computation of Fourier pairs. We emphasise that BRFP, in addition to being a method for spectral/temporal reconstruction, is a generative model for both temporal and spectral objects, and as such allows for a joint reconstruction in both domains, while being competitive with purpose-specific methods as validated in the experimental section. Furthermore, the superior performance of BRFP is supported by the fact that the method integrates over the latent quantities, and although BNSE \cite{TobarNIPS2018} also has this property, BRFP does it with less structural (path integrability) assumptions. 

The theoretical and practical contributions position BRFP as a unique yet simple methodology for modelling Fourier pairs, while also providing an account of its own uncertainty and therefore addressing the question \emph{how trustworthy is our reconstruction?} This is critical for decision-making and designing new hardware for spectral acquisition.

Building on the concepts in this work, future research directions  include, in the Gaussian case, the design of novel covariance structures for BRFP, which encode specific time-series properties such as quasi-periodicity, differentiability, and seasonality. The main research direction, however, should be towards relaxing the Gaussian assumption on the time series, though this is challenging due to the conjugacy between the Gaussian distribution and the Fourier operator.} One way of lifting this restriction would be to consider a latent representation of the data, which could be Gaussian as in \cite{rios19}, so that the observed data are a (nonlinear) transformation of such a hidden representation. This is equivalent to considering a nonlinear likelihood $\Ha$ in the proposed model. A possibility for such a nonlinearity is to consider fuzzy approximations as in \cite{8957082,9026974}.

\section*{Acknowledgments}
The authors thank financial support form the following grants: Fondecyt-Iniciación 11171165, ANID-AFB170001 Center for Mathematical Modeling, ANID-FB0008 Center for Electrical and Electronic Engineering, ANID Millennium Initiative IC12009 Millennium Institute of Astrophysics, ANID PFCHA/MagísterNacional/2017-22171830, and the Growing Convergence Research Program of the National Science Foundation: Award number 2021002.

This paper makes use of the following ALMA data: ALMA\#2011.0.00465.S. ALMA is a partnership of ESO, NSF, NINS, NRC, NSC, ASIAA. The Joint ALMA Observatory is operated by ESO, AUI/NRAO, and NAOJ. We are thanful to Axel Osses, Pablo Román and Jorge Silva for discussions and advice on how to approach this dataset.

\bibliographystyle{IEEEbib}
\bibliography{references}

\begin{thebibliography}{10}

\bibitem{NRAO}
G.~B. {Taylor}, C.~L. {Carilli}, and R.~A. {Perley}, Eds.,
\newblock {\em {Synthesis Imaging in Radio Astronomy II}}, vol. 180 of {\em
  Astronomical Society of the Pacific Conference Series}, 1999.

\bibitem{Stoica2005}
P.~Stoica and M.~Randolph,
\newblock {\em Spectral Analysis of Signals},
\newblock Prentice Hall, Inc., 2005.

\bibitem{kay:88}
S.~Kay,
\newblock {\em Modern {S}pectral {E}stimation : {T}heory and {A}pplication},
\newblock Englewood Cliffs, N.J. : Prentice Hall, 1988.

\bibitem{Welch1967}
P.~Welch,
\newblock ``The use of fast {F}ourier transform for the estimation of power
  spectra: A method based on time averaging over short, modified
  periodograms,''
\newblock {\em IEEE Transactions on Audio and Electroacoustics}, vol. 15, no.
  2, pp. 70--73, 1967.

\bibitem{1965-cooley}
J.~W. Cooley and J.~W. Tukey,
\newblock ``{An algorithm for the machine calculation of complex Fourier
  series},''
\newblock {\em Mathematics of Computation}, vol. 19, pp. 297--301, 1965.

\bibitem{Yule267}
G.~U. Yule,
\newblock ``On a method of investigating periodicities in disturbed series,
  with special reference to {W}olfer{\textquoteright}s sunspot numbers,''
\newblock {\em Phil.~Trans.~of the Royal Society of London A: Mathematical,
  Physical and Engineering Sciences}, vol. 226, no. 636-646, pp. 267--298,
  1927.

\bibitem{Walker518}
G.~Walker,
\newblock ``On periodicity in series of related terms,''
\newblock {\em Proc. of the Royal Society of London A: Mathematical, Physical
  and Engineering Sciences}, vol. 131, no. 818, pp. 518--532, 1931.

\bibitem{Lomb1976}
N.~R. Lomb,
\newblock ``Least-squares frequency analysis of unequally spaced data,''
\newblock {\em Astrophysics and Space Science}, vol. 39, no. 2, pp. 447--462,
  feb 1976.

\bibitem{Scargle1982}
J.~D. Scargle,
\newblock ``Studies in astronomical time series analysis. {II} - statistical
  aspects of spectral analysis of unevenly spaced data,''
\newblock {\em The Astrophysical Journal}, vol. 263, pp. 835, dec 1982.

\bibitem{jaynes1987bayesian}
E.~T. Jaynes,
\newblock ``Bayesian spectrum and chirp analysis,''
\newblock in {\em Maximum-Entropy and Bayesian Spectral Analysis and Estimation
  Problems}, pp. 1--37. Springer, 1987.

\bibitem{bretthorst2013bayesian}
G.~L. Bretthorst,
\newblock {\em Bayesian Spectrum Analysis and Parameter Estimation},
\newblock Lecture Notes in Statistics. Springer, 1988.

\bibitem{qi_minka_picard}
Y.~Qi, T.~P. Minka, and R.~W. Picard,
\newblock ``Bayesian spectrum estimation of unevenly sampled nonstationary
  data,''
\newblock in {\em Proc. of ICASSP}, 2002, vol.~2, pp. 1473--1476.

\bibitem{TobarNIPS2018}
F.~Tobar,
\newblock ``Bayesian nonparametric spectral estimation,''
\newblock in {\em Advances in Neural Information Processing Systems},
  S.~Bengio, H.~Wallach, H.~Larochelle, K.~Grauman, N.~Cesa-Bianchi, and
  R.~Garnett, Eds. 2018, vol.~31, pp. 10127--10137, Curran Associates, Inc.

\bibitem{protopapas}
Y.~Wang, R.~Khardon, and P.~Protopapas,
\newblock ``Nonparametric {B}ayesian estimation of periodic light curves,''
\newblock {\em The Astrophysical Journal}, vol. 756, no. 1, pp. 67, 2012.

\bibitem{hall2006resolution}
M.~Hall,
\newblock ``Resolution and uncertainty in spectral decomposition,''
\newblock {\em First Break}, vol. 24, no. 12, 2006.

\bibitem{Murphy2012}
K.~P. Murphy,
\newblock {\em Machine Learning: A Probabilistic Perspective},
\newblock The MIT Press, 2012.

\bibitem{Rasmussen:2005:GPM:1162254}
C.~E. Rasmussen and C.~K.~I. Williams,
\newblock {\em Gaussian Processes for Machine Learning},
\newblock The MIT Press, 2005.

\bibitem{mackay_gp}
D.~J.~C. MacKay,
\newblock ``Introduction to {G}aussian processes,''
\newblock {\em NATO ASI Series F Computer and Systems Sciences}, vol. 168, pp.
  133--166, 1998.

\bibitem{TobarTuner2015}
F.~Tobar, T.~D. Bui, and R.~E. Turner,
\newblock ``Learning stationary time series using {G}aussian processes with
  nonparametric kernels,''
\newblock in {\em Advances in Neural Information Processing Systems 28}, pp.
  3501--3509. Curran Associates, Inc., 2015.

\bibitem{nips17}
G.~Parra and F.~Tobar,
\newblock ``Spectral mixture kernels for multi-output {G}aussian processes,''
\newblock in {\em Advances in Neural Information Processing Systems 30}, pp.
  6681--6690. Curran Associates, Inc., 2017.

\bibitem{tobar19b}
F.~Tobar,
\newblock ``Band-limited {G}aussian processes: The sinc kernel,''
\newblock in {\em Advances in Neural Information Processing Systems 32}, pp.
  12749--12759. Curran Associates, Inc., 2019.

\bibitem{tobar18a}
B.~Poblete, J.~Guzman, J.~Maldonado, and F.~Tobar,
\newblock ``Robust detection of extreme events using {T}witter: {W}orldwide
  earthquake monitoring,''
\newblock {\em IEEE Transactions on Multimedia}, vol. 20, no. 10, pp.
  2551--2561, 2018.

\bibitem{tobar20a}
T.~de~Wolff, A.~Cuevas, and F.~Tobar,
\newblock ``Gaussian process imputation of multiple financial series,''
\newblock in {\em Proc. of IEEE ICASSP}, 2020, pp. 8444--8448.

\bibitem{ijcnn18}
G.~Rios and F.~Tobar,
\newblock ``Learning non-{G}aussian time series using the {B}ox-{C}ox
  {G}aussian process,''
\newblock in {\em Proc. of the International Joint Conference on Neural
  Networks}, 2018, pp. 1--8.

\bibitem{rios19}
G.~Rios and F.~Tobar,
\newblock ``Compositionally-warped {G}aussian processes,''
\newblock {\em Neural Networks}, vol. 118, pp. 235 -- 246, 2019.

\bibitem{warped04}
E.~Snelson, Z.~Ghahramani, and C.~E. Rasmussen,
\newblock ``Warped {G}aussian processes,''
\newblock in {\em Advances in Neural Information Processing Systems 16}, pp.
  337--344. MIT Press, 2004.

\bibitem{bayesianwarped12}
M.~L\'{a}zaro-Gredilla,
\newblock ``Bayesian warped {G}aussian processes,''
\newblock in {\em Advances in Neural Information Processing Systems 25}, pp.
  1619--1627. Curran Associates, Inc., 2012.

\bibitem{Rao2000}
K.~R. Rao and P.~C. Yip, Eds.,
\newblock {\em The Transform and Data Compression Handbook},
\newblock CRC Press, Inc., Boca Raton, FL, USA, 2000.

\bibitem{MandicGoh}
D.~P. Mandic and V.~S.~L. Goh,
\newblock {\em Complex Valued Nonlinear Adaptive Filters: Noncircularity,
  Widely Linear and Neural Models},
\newblock Wiley, 2009.

\bibitem{icassp15}
F.~Tobar and R.~Turner,
\newblock ``Modelling of complex signals using {G}aussian processes,''
\newblock in {\em Proc. of IEEE ICASSP}, 2015, pp. 2209--2213.

\bibitem{flamary2016optimal}
R.~Flamary, C.~F{\'e}votte, N.~Courty, and V.~Emiya,
\newblock ``Optimal spectral transportation with application to music
  transcription,''
\newblock in {\em Advances in Neural Information Processing Systems}. 2016, pp.
  703--711, Curran Associates, Inc.

\bibitem{Wasserstein-Fourier}
E.~{Cazelles}, A.~{Robert}, and F.~{Tobar},
\newblock ``The {W}asserstein-{F}ourier distance for stationary time series,''
\newblock {\em arXiv e-prints}, p. arXiv:1912.05509, 2019.

\bibitem{Goldberger1991}
A.~L. Goldberger and D.~R. Rigney,
\newblock {\em Nonlinear Dynamics at the Bedside}, pp. 583--605,
\newblock Springer New York, 1991.

\bibitem{Thompson2017}
A.~R. Thompson, J.~M. Moran, and G.~W. Swenson,
\newblock {\em Introductory Theory of Interferometry and Synthesis Imaging},
  pp. 59--88,
\newblock Springer International Publishing, Cham, 2017.

\bibitem{CLEAN}
J.~A. Högbom,
\newblock ``Aperture synthesis with a non-regular distribution of
  interferometer baselines,''
\newblock {\em Astronomy and Astrophysics Supplement}, vol. 15, pp. 417--426,
  1974.

\bibitem{Casassus2015accretion}
S.~Casassus, S.~Marino, S.~P{\'e}rez, P.~Roman, A.~Dunhill, P.~J. Armitage,
  J.~Cuadra, A.~Wootten, G.~van~der Plas, L.~Cieza, et~al.,
\newblock ``Accretion kinematics through the warped transition disk in {HD}
  142527 from resolved co (6--5) observations,''
\newblock {\em The Astrophysical Journal}, vol. 811, no. 2, pp. 92, 2015.

\bibitem{8957082}
K.~{Sun}, L.~{Liu}, J.~{Qiu}, and G.~{Feng},
\newblock ``Fuzzy adaptive finite-time fault-tolerant control for
  strict-feedback nonlinear systems,''
\newblock {\em IEEE Transactions on Fuzzy Systems}, 2020,
\newblock in press, doi: 10.1109/TFUZZ.2020.2965890.

\bibitem{9026974}
K.~{Sun}, J.~{Qiu}, H.~R. {Karimi}, and Y.~{Fu},
\newblock ``Event-triggered robust fuzzy adaptive finite-time control of
  nonlinear systems with prescribed performance,''
\newblock {\em IEEE Transactions on Fuzzy Systems}, 2020,
\newblock in press, doi: 10.1109/TFUZZ.2020.2979129.

\bibitem{GLS}
M.~Zechmeister and M.~K\"urster,
\newblock ``The generalised {Lomb-Scargle} periodogram. a new formalism for the
  floating-mean and {K}eplerian periodograms,''
\newblock {\em A\&A}, vol. 496, no. 2, pp. 577--584, 2009.

\bibitem{BGLS}
A.~Mortier, J.P. Faria, C.~M. Correia, A.~Santerne, and N.~C. Santos,
\newblock ``{BGLS}: A {B}ayesian formalism for the generalised {Lomb-Scargle}
  periodogram,''
\newblock {\em A\&A}, vol. 573, pp. A101, 2015.

\bibitem{Cands2006}
E.~J. Cand{\`{e}}s, J.~K. Romberg, and T.~Tao,
\newblock ``Stable signal recovery from incomplete and inaccurate
  measurements,''
\newblock {\em Communications on Pure and Applied Mathematics}, vol. 59, no. 8,
  pp. 1207--1223, 2006.

\bibitem{Cands2008}
E.~J. Cand{\`{e}}s,
\newblock ``The restricted isometry property and its implications for
  compressed sensing,''
\newblock {\em Comptes Rendus Mathematique}, vol. 346, no. 9-10, pp. 589--592,
  May 2008.

\bibitem{Foucart2013}
S.~Foucart and H.~Rauhut,
\newblock {\em A Mathematical Introduction to Compressive Sensing},
\newblock Springer New York, 2013.

\end{thebibliography}
%!TEX root = ./BRFP.tex

\appendix

\subsection{Benchmarks}
\label{sec:benchmarks}

\revised{Experiments E1, E2 and E3 (Secs.~ \ref{sec:E1}, \ref{sec:E2} and \ref{sec:simulations_alma} respectively) aim to validate the proposed BRFP against various methods for spectral estimation. We briefly describe those benchmarks here.\\
$\bullet$ \textbf{YW} (Yule-Walker, \cite{Yule267,Walker518}): Modelling the data as an autoregressive (AR) process and learning the weights via least squares. Estimate PSD is that of the learnt AR model. Code from \emph{pyspectrum\footnote{Source: \url{https://pyspectrum.readthedocs.io/en/latest/}}}.\\
\noindent $\bullet$ \textbf{LS} (Lomb-Scargle, \cite{Lomb1976,Scargle1982}): A sum-of-sinusoid model is fitted to the data via least squares. Estimate PSD is that of the learnt model. Code from \emph{astropy}\footnote{Source: \url{https://www.astropy.org}}.\\
\noindent $\bullet$ \textbf{Welch} (Welch Periodogram, \cite{Welch1967}): The data is split into batches, and the standard Periodogram is computed for each batch. The estimate PSD is the (Euclidean) average of all per-batch periodograms. Code from \emph{scipy\footnote{Source: \url{https://www.scipy.org}}}\\
\noindent $\bullet$ \textbf{BGLS} (Bayesian Generalised LS, \cite{GLS,BGLS}): A variant of the LS method above that includes a floating mean and a Gaussian noise model. Code from \cite{BGLS}.\\
\noindent $\bullet$ \textbf{CS-DCT} (Compressed Sensing, \cite{Cands2006, Cands2008, Foucart2013}): Sparse reconstruction of the data using the the discrete cosine transform (DCT) and the LASSO ($\ell_1$) cost. The estimate PSD is given by DCT representation. Code from \emph{pyrunner}\footnote{Source: \url{http://www.pyrunner.com/weblog/2016/05/26/compressed-sensing-python/}}.\\
\noindent $\bullet$ \textbf{BNSE} (Bayesian Nonparametric Spectral Estimation, \cite{TobarNIPS2018}): The observations are interpolated by a Gaussian process (GP). The estimate PSD is the Fourier transform of the (windowed) posterior GP mean. Code from \cite{TobarNIPS2018}.

We clarify that not all these methods are suitable for every experiment due to their natural limitations, e.g., the requirement for evenly-spaced data. Therefore, only subsets of the above list were used in some experiments.
}

\subsection{Model formulation for 2D signals}
\label{sec:2dmodel}
For completeness, and in particular to clarify how experiment \ref{sec:simulations_alma} was implemented, we briefly explain how the proposed generative model can deal with bi-dimensional data, or \emph{images}. 

Our strategy was to \emph{vectorise} the 2D data, converting it into 1D and thus compatible with the proposed BRFP. To this end, we considered an image $\mathbf{i} \in \mathbb{R}^{N\times N}$ and its corresponding  Fourier transform $\mathbf{I} \in \mathbb{C}^{N\times N}$; these two objects are 2D Fourier pairs related by:
\begin{equation}
    \mathbf{I} = W^\top \mathbf{i} W, \label{eq:imgFourier}
\end{equation}
where $W$ is the Fourier matrix defined in eq.~\eqref{eq:fourierOp}. The above expression can be rewritten in vectorised form, or applying the \emph{unfolding} operator. This required to replace the matrix product by the Kronecker product to yield
\begin{align}
\mathbf{I} & = W^\top \mathbf{i} W & / \text{vec}(\cdot),\\
\vec{\mathbf{I}} & = (W^\top \otimes W^\top)\vec{\mathbf{i}},\\
\vec{\mathbf{I}} & = \mathcal{W}^\top\vec{\mathbf{i}}.
\end{align}
This last expression is of the form of two 1D Fourier pairs, where the operator is given by the matrix $\mathcal{W}=W^\top \otimes W^\top$. Therefore, $\vec{\mathbf{i} }$-$\vec{\mathbf{I}}$ follows a relationship of the form assumed in the paper, therefore, we can apply BRFP.

We clarify that the design of the covariance $\Sigma$ in eq.~\eqref{eq:time_series} needs to take into account the vectorised representation of the image. We addressed this by designing the covariance for the image representation $\mathbf{i}$, and then \emph{unfolding} the kernel as well, so that it corresponds to the kernel of the vectorised image $\vec{\mathbf{I}}$. As a consequence, we have a complete vector representation of the image, including its covariance matrix, that is compatible with the original formulation of BRFP.

\end{document}